\documentclass[prb, twocolumn, superscriptaddress, showpacs]{revtex4-2}
\usepackage[english]{babel}
\usepackage{hyperref}
\usepackage{physics}
\usepackage{dsfont}
\usepackage{amsfonts}
\usepackage{bm}
\usepackage{graphicx}
\usepackage{xcolor}
\usepackage{mathtools}
\usepackage{verbatim}

\begin{document}

\title{Logarithmic entanglement scaling in dissipative free-fermion systems}
\date{\today}
\author{Antonio D'Abbruzzo}
\affiliation{Scuola Normale Superiore, I-56126 Pisa, Italy}
\author{Vincenzo Alba}
\affiliation{Dipartimento di Fisica dell'Universit\`a di Pisa
and INFN, Sezione di Pisa, I-56127 Pisa, Italy}
\author{Davide Rossini}
\affiliation{Dipartimento di Fisica dell'Universit\`a di Pisa
and INFN, Sezione di Pisa, I-56127 Pisa, Italy}

\begin{abstract}
	We study the quantum information spreading in one-dimensional free-fermion 
	systems in the presence of localized thermal baths.
	We employ a \emph{nonlocal} Lindblad master equation to describe the 
	system-bath interaction, in the sense that the Lindblad operators are 
	written in terms of the Bogoliubov operators of the closed system, and 
	hence are nonlocal in space.
	The statistical ensemble describing the steady state is written in terms of 
	a convex combination of the Fermi-Dirac distributions of the baths.
	Due to the singularity of the free-fermion dispersion, the steady-state 
	mutual information exhibits singularities as a function of the system 
	parameters.
	While the mutual information generically satisfies an area law, at the 
	singular points it exhibits logarithmic scaling as a function of subsystem 
	size. By employing the Fisher-Hartwig theorem, we derive the 
	prefactor of the logarithmic scaling, which depends on the parameters of 
	the baths and plays the role of an effective ``central charge''. This is 
	upper bounded by the central charge governing ground-state entanglement 
	scaling.
	We provide numerical checks of our results in the paradigmatic 
	tight-binding chain and the Kitaev chain.
\end{abstract}

\maketitle


\section{Introduction} \label{sec:intro}

The study of the interplay between the microscopic quantum world and the 
macroscopic classical one is a fundamental research topic in contemporary 
physics, although it dates back to the first days of quantum
mechanics~\cite{Zurek_2003, rossini2021coherent}. Typically, the interaction 
with the environment is believed to destroy genuine quantum behaviors, although 
consensus is emerging that this is not always the case.
Dissipation-based protocols have been devised to imprint nontrivial 
correlations in quantum many-body systems
(see, e.g., Refs.~\cite{Syassen_2008, lin2013, diehl_2008, verstraete-2009, 
Eisert_2010, Roncaglia_2010, diehl-2011, Bouchoule_2020, Rossini_2021, 
seetharam2022correlation}).
A first crucial question is whether entanglement, which is the distinctive 
feature of quantum mechanics, is robust against the presence of the 
environment. Second, is it possible to enhance the entanglement content of a 
quantum many-body state via an \emph{ad hoc} engineered environment?
Answering these questions is a daunting task, because there is no universal 
approach (neither analytic nor numerical) to tackle generic \emph{open} quantum 
many-body systems. With this state of affairs, one has to resort to approximate 
treatments. Markovian master equations, such as the Lindblad master 
equation~\cite{lindblad1976on, gorini1976completely, petruccione2002the},
provide some of the most successful tools to attack open quantum many-body 
systems. 

\begin{figure}
	\centering
	\includegraphics[width=\columnwidth]{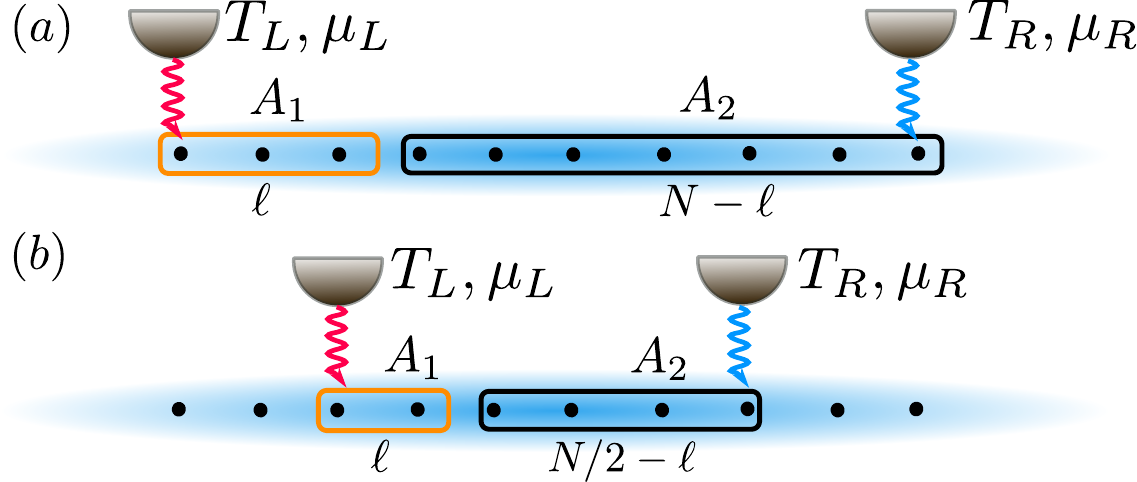}
	\caption{
		The setup used in this work. A one-dimensional chain of $N$ 
		noninteracting fermions is put in contact with two thermal reservoirs 
		at temperatures $T_L$, $T_R$ and chemical potentials $\mu_L$, $\mu_R$. 
		(a) In the case of OBC, only the sites at the edges are in contact with 
		the baths, and we are interested in the steady-state mutual information
		$I(A_1:A_2)$ between the two connected intervals $A_1$ and $A_2$ of 
		length $\ell$ and $N-\ell$, respectively.
		(b) In the case of PBC, the sites in contact with the baths are at the 
		maximum distance $N/2$ and the intervals $A_1$ and $A_2$ are 
		respectively of length $\ell$ and $N/2-\ell$. Analytical calculations 
		are done in the thermodynamic limit $N,\ell \to \infty$.
	}
	\label{fig:sb}
\end{figure}

Particularly important settings are provided by the class of nonequilibrium
boundary-driven quantum systems, which have been the subject of intense
research in recent years (see, e.g., Refs.~\cite{bertini2021,Landi2021}
and references therein).
In this paper we focus on the one-dimensional setup illustrated in 
Fig.~\ref{fig:sb}: a system of \emph{noninteracting} fermions is locally 
coupled to ideal thermal baths. 
To be specific, we focus on the tight-binding chain and on the Kitaev chain. 
The fermions live on a lattice with $N$ sites, with either periodic boundary 
conditions (PBC) or open boundary conditions (OBC).
The system is put in contact with two ideal fermionic reservoirs at 
temperatures $T_L,T_R$, and with chemical potentials $\mu_L,\mu_R$. With OBC 
the two baths are placed at the edges of the chain [Fig.~\ref{fig:sb}(a)], 
whereas with PBC they are are at the maximum distance $N/2$
[Fig.~\ref{fig:sb}(b)].
The interaction between the chain and the reservoirs is treated within the 
formalism of the Lindblad master equation~\cite{petruccione2002the}. 
Specifically, we employ the \emph{nonlocal} description derived in
Ref.~\cite{dabbruzzo2021self}.
The Lindblad operators are obtained \emph{ab initio} from the microscopic 
system-bath interaction, and are written in terms of the Bogoliubov modes of 
the model without dissipation. As such, the Lindblad operators are non-local in 
real space. 
Interestingly, this allows to recover the Conformal Field Theory (CFT) 
description of the chain in the low-temperature limit.
Furthermore, the nonlocal Lindblad approach allows to obtain a 
thermodynamically consistent description of transport
properties~\cite{dabbruzzo2021self, dabbruzzo2021self2}. 

We are interested in the quantum correlations emerging in the steady state of 
finite chains of length $N$, in the limit $t \to \infty$. The ensemble 
describing this state is written in terms of a convex combination of the 
Fermi-Dirac distributions of the reservoirs~\cite{dabbruzzo2021self}.
Importantly, this ensemble is in general different from the finite-temperature 
ensemble of the underlying fermionic chain. In particular, we will show that 
ground-state criticality of the free chain Hamiltonian is associated with 
nontrivial steady-state correlations in the dissipative model.
To monitor these correlations, we consider the quantum mutual information
$I(A_1:A_2)$ between two subregions $A_1$ and $A_2$ of the chain, defined
as~\cite{amico2008entanglement, calabrese2009entanglemententropy, 
eisert2010colloquium, laflorencie2016quantum}
\begin{equation} \label{eq:mi-intro}
	I(A_1:A_2) \coloneqq S_{A_1} + S_{A_2} - S_{A_1\cup A_2}, 
\end{equation}
where $S_A$ is the von Neumann entropy of the subregion $A$,
which is defined as 
\begin{equation} \label{eq:vn}
	S_A \coloneqq -\Tr[\rho_A \ln \rho_A], 
\end{equation}
where $\rho_A$ is the reduced density matrix for subsystem $A$.

For pure states the von Neumann entropy $S_A$ of a subsystem quantifies its 
entanglement with the rest of the system. Moreover, one has that
$S_A = S_{\overline{A}}$, with $\overline{A}$ being the complement of $A$
[see, for instance, Fig.~\ref{fig:sb}(a)], and $S_{A\cup\overline{A}}=0$. 
However, in the presence of an environment, the global state is mixed, which 
implies that neither the von Neumann entropy nor the mutual information are 
proper measures of the entanglement shared between different regions. 
Still, it has been shown recently that for out-of-equilibrium free-fermion and 
free-boson models in the presence of quadratic global dissipation the mutual 
information admits a hydrodynamic description in terms of a quasiparticle 
picture~\cite{alba2021spreading,carollo2022dissipative,alba2022hydrodynamics,
alba2022logarithmic}.
Moreover, we numerically checked that our findings remain qualitatively valid 
using the fermionic entanglement negativity, which, on the other hand, is
a proper measure of entanglement~\cite{shapourian2017,shapourian2019}.

Both the full-system entropy, as well as the subsystems entropies, generally 
exhibit volume-law scaling in the steady state.  
However, if the bulk of the system is tuned to a critical point, logarithmic 
corrections appear. Specifically we show that, for a generic subsystem $A$
of length $\ell$, in the scaling limit $N,\ell\to\infty$ with arbitrary ratio 
$\ell/N$, the entropy $S_A$ is given as 
\begin{equation}\label{eq:ent-asy}
	S_A = \alpha \ell+\frac{c(\Theta)}{3\nu}\ln\left[\frac{N}{\pi}\sin\left(
	\frac{\pi\ell}{N}\right)\right]+{\mathcal O}(1). 
\end{equation}
The prefactor $\alpha$ of the volume-law term depends on the full spectrum of 
the model and on the properties of the bath. 
A similar volume-law scaling has been found in free-fermion systems with 
localized dissipative
impurities~\cite{alba2022noninteracting,alba2022unbounded}. 
The prefactor $c(\Theta)$ is an effective ``central charge'', which contains 
information only about the singularities in the single-particle spectrum of the 
model. It is an even function of a parameter $\Theta \in [-1,1]$ which depends 
on the properties of the baths, i.e., their temperatures and chemical  
potentials.
The quantity $\nu \in \{1,2\}$ depends on the considered setting:
for the situation in Fig.~\ref{fig:sb}(a) with OBC we have $\nu=2$,
while for the case in Fig.~\ref{fig:sb}(b) with PBC we have $\nu = 1$.
The argument inside the square brackets is the so-called
chord length~\cite{difrancesco1997conformal}. 
In the limit $\ell/N\to 0$, the second term in~\eqref{eq:ent-asy} becomes 
$c/(3\nu)\ln(\ell)$, which we prove analytically.
On the other hand, the result for finite ratio $\ell/N$ is a conjecture 
inspired by the
zero-temperature CFT scaling~\cite{calabrese2009entanglemententropy}. 
Similar logarithmic terms as in~\eqref{eq:ent-asy} have been found in a 
tight-binding model, although for different nonequilibrium
settings~\cite{Eisler2014, Kormos2017, fraenkel2021, turkeshi2022entanglement,
turkeshi2022enhanced}. 
Finally, the last term ${\mathcal O}(1)$ is a subleading constant, which can be 
calculated, at least for the tight-binding chain, by using the Fisher-Hartwig 
conjecture~\cite{fisher1969toeplitz,basor1991the,basor1994the,
forrester2004applications,deifts2011asymptotics,fagotti2011universal}. 

In the limit $|\Theta|\to1$ one recovers the zero-temperature result, i.e.,
$c(\Theta)$ becomes the central charge of the CFT that describes ground-state 
properties of the model. Here one has $c=1$ and $c=1/2$ for the
tight-binding chain and the Kitaev chain, respectively.
On the other hand, for $\Theta\to0$, which corresponds to the high-temperature
limit, $c(\Theta)$ vanishes. Remarkably, for generic $\Theta$, the effective
central charge of the tight-binding chain is twice that of the Kitaev model.
Moreover, we show that $c(\Theta)$ is always upper bounded by the 
zero-temperature central charge of the models. 
Upon substituting the asymptotic scaling~\eqref{eq:ent-asy}
in~\eqref{eq:mi-intro}, one obtains that the volume-law term cancels out,
and the mutual information exhibits a logarithmic scaling. 

The structure of the paper is as follows.
In Sec.~\ref{sec:models} we introduce the fermionic models we are interested in.
Sec.~\ref{sec:majo} contains the calculation of the Majorana correlation matrix, 
which is important to determine entanglement properties, while the main formulas
to determine the von Neumann entropies in terms of correlation functions
are reviewed in Sec.~\ref{sec:ent-free-f}. 
In Sec.~\ref{sec:lindblad} we summarize the treatment of thermal environments
within the nonlocal Lindblad equation, which was derived in
Ref.~\cite{dabbruzzo2021self}; the main result is formula~\eqref{eq:statio}.
In Sec.~\ref{sec:tb-ent} and Sec.~\ref{sec:kitaev-ent} we analytically
derive Eq.~\eqref{eq:ent-asy} for the tight-binding chain and for the
Kitaev chain, respectively, in the limit $\ell/N \to 0$.
The details of the computations are deferred to App.~\ref{app:tight-binding}
and App.~\ref{app:kitaev}, respectively.
We provide numerical benchmarks of our results in Sec.~\ref{sec:numerics}:
in particular, in Sec.~\ref{sec:vn} we overview the volume-law scaling of the
von Neumann entropy, in Sec.~\ref{sec:mi} we discuss the scaling of the 
mutual information, and in Sec.~\ref{sec:negativity} we briefly
discuss the behavior of the fermionic logarithmic negativity.
Our conclusions are drawn in Sec.~\ref{sec:conclusions}.
Appendix~\ref{app:fourier} contains a proof that our results are not
an artifact dictated by the choice of the basis adopted
to diagonalize the model.


\section{Models \& methods} \label{sec:models}

In this section we describe the basic framework used in this work.
We first introduce the quadratic fermionic Hamiltonians of interest
in Sec.~\ref{sec:free-fermions}. 
Then we review the tight-binding chain (Sec.~\ref{sec:tb})
and the Kitaev chain (Sec.~\ref{sec:kitaev}). 
In Sec.~\ref{sec:majo} we provide some general formulas for the 
two-point correlation of Majorana operators in the two models.
These are essential to study entanglement properties. 
In Sec.~\ref{sec:ent-free-f} we summarize the calculation of the von Neumann 
entropy via correlation matrix techniques~\cite{jin2004quantum}. 
In Sec.~\ref{sec:lindblad} we review the approach of
Ref.~\cite{dabbruzzo2021self} to derive a \emph{nonlocal} Lindblad description
of fermionic chains in contact with localized thermal baths.


\subsection{Fermionic quadratic Hamiltonians} \label{sec:free-fermions}

We focus on free-fermion chains~\cite{lieb1961two, pfeuty1970the}. 
Let $\mathcal{S}$ be a quantum system on a lattice with $N$ sites
and $H_\mathcal{S}$ its second-quantized Hamiltonian, which
can be written in terms of fermionic raising and lowering operators
$a_n^\dagger, a_n$, with $n \in \{1,\ldots,N\}$.
We assume $H_\mathcal{S}$ to be quadratic, i.e.,
\begin{equation} \label{eq:ham-gen}
	H_\mathcal{S} = \sum_{n,m = 1}^N \left[ Q_{nm} a_n^\dagger a_m +
	\frac{P_{nm}}{2} \left( a_n^\dagger a_m^\dagger - a_n a_m \right) \right],
\end{equation}
with $Q,P$ being $N \times N$ real matrices satisfying $Q^T = Q$ and $P^T = -P$.
It is known that $H_\mathcal{S}$ can be written in diagonal form as
\begin{equation} \label{eq:ek}
	H_\mathcal{S} = E_0 + \sum_k \omega_k b_k^\dagger b_k \, ,
\end{equation}
where $\omega_k \geq 0$ is the single-particle dispersion, 
$E_0$ is an irrelevant constant, $b_k^\dagger, b_k$ are new fermionic operators
(the Bogoliubov modes), and the index $k$ denotes the quasimomentum. 
The operators $b_k$ are written as linear superpositions of the original
fermions. Specifically, one has 
\begin{equation} \label{eq:bk-def}
	b_k = \sum_{n=1}^N \left( X_{kn} a_n + Y_{kn}a_n^\dagger \right),
\end{equation}
where $X$ and $Y$ are appropriately chosen $N \times N$ complex matrices.
For the following it is useful to define the so-called
Lieb-Schultz-Mattis matrices~\cite{lieb1961two} $\phi$, $\psi$ as 
\begin{equation} \label{eq:phipsi-def}
	\phi \coloneqq (X+Y)^\dagger, \qquad \psi \coloneqq (X-Y)^\dagger.
\end{equation}
In general, these are complex $N \times N$ matrices that encode
information about the Majorana correlation functions of the model
(see Sec.~\ref{sec:majo}).


\subsubsection{Tight-binding chain} \label{sec:tb}

The tight-binding model is obtained from~\eqref{eq:ham-gen} with 
\begin{equation}
	Q_{nm}=-J\delta_{n,m-1}-J\delta_{n,m+1}-h\delta_{nm},
	\quad P_{nm}=0,
\end{equation}
where $h$ is an external magnetic field strength, and $J$ is the 
hopping amplitude between nearest-neighbor sites. 
Thus, the Hamiltonian reads as
\begin{equation} \label{eq:ham-xx}
	H_\mathcal{S} = -J\sum_{n=1}^{N} \left( a_n^\dagger a_{n+1} +
	a_{n+1}^\dagger a_n \right) - h \sum_{n=1}^N a_n^\dagger a_n,
\end{equation}
where $a_{N+1}$ is determined by the boundary conditions: with OBC
we are choosing $a_{N+1} = 0$, whereas with PBC we have $a_{N+1} = a_1$.
For simplicity, hereafter we set $J=1$ and work in units of
$\hbar = k_B = 1$.

The single-particle dispersion relation [cf.~Eq.~\eqref{eq:ek}] is given by
\begin{equation} \label{eq:xx-obc-disp}
	\omega_k = \big| h + 2\cos k \big|, 
\end{equation}
where ($n=1,\ldots,N$):
\begin{equation} \label{eq:k_bc}
	k = \begin{cases}
		n\pi / (N+1) & \text{(OBC)}, \\
		2\pi n/N & \text{(PBC)}.
	\end{cases}
\end{equation}
The functions $\phi_{nk}$ and $\psi_{nk}$ [cf.~Eq.~\eqref{eq:phipsi-def}]
are given by 
\begin{subequations} \label{eq:phipsi-xx}
	\begin{align}
		\label{eq:phi-xx} \phi_{nk} &= \begin{cases}
			\sqrt{2/(N+1)} \sin(kn) & \text{(OBC)}, \\
			e^{ikn}/\sqrt{N} & \text{(PBC)},
		\end{cases} \\
		\label{eq:psi-xx} \psi_{nk} &= \mathrm{sgn}( -h-2\cos k ) \phi_{nk},
	\end{align}
\end{subequations}
where $\mathrm{sgn}(x)$ is the sign function and
the corresponding quasimomentum index has to be chosen as in~\eqref{eq:k_bc}.
The ground state of the tight-binding model is annihilated by 
all the Bogoliubov operators $b_k$ [cf.~Eq.~\eqref{eq:bk-def}], 
and it exhibits criticality
in the conducting phase $|h| \leq 2$, where
its properties are described
by a CFT~\cite{difrancesco1997conformal} with central charge $c=1$. 

The entanglement properties of free-fermion systems are encoded 
in the fermionic two-point correlation functions~\cite{peschel2009reduced}. 
Let us first discuss the tight-binding chain with OBC.
In the limit $N\to\infty$, the ground-state fermionic correlation function 
$C^{(\mathrm{obc})}_{nm} \coloneqq \langle a^\dagger_n a_m\rangle$
is given as~\cite{peschel2009reduced}  
\begin{equation} \label{eq:gs-corr}
	C^{(\mathrm{obc})}_{nm} = \int_{-\pi}^\pi \frac{dk}{2\pi} \,
	\Theta_{\mathrm{H}}(k_F-|k|) \left[ e^{ik(n-m)} - e^{ik(n+m)} \right], 
\end{equation}
where $\Theta_{\mathrm{H}}(\cdot)$ is the Heaviside step function, 
and $k_F$ is the Fermi momentum
\begin{equation} \label{eq:kf}
	k_F \coloneqq \arccos\left(-\frac{h}{2}\right). 
\end{equation}
Performing the integral in~\eqref{eq:gs-corr}, one obtains 
\begin{equation} \label{eq:gs-corr-1}
	C^{(\mathrm{obc})}_{nm}=\frac{\sin(k_F(n-m))}{\pi(n-m)}-
	\frac{\sin(k_F(n+m))}{\pi(n+m)}.
\end{equation}
The result for the infinite chain with PBC can be recovered
from~\eqref{eq:gs-corr-1} by taking the limit $n,m\to\infty$ with
$n-m$ fixed, i.e., by considering correlators in the bulk of the open chain. 
Thus, only the first term in~\eqref{eq:gs-corr} survives and one obtains 
\begin{equation} \label{eq:gs-corr-pbc}
	C^{(\mathrm{pbc})}_{nm} = \frac{\sin(k_F(n-m))}{\pi(n-m)}. 
\end{equation}
It is useful to observe that the first term in~\eqref{eq:gs-corr} 
depends only on the difference $n-m$, reflecting translation invariance, and it
defines a so-called Toeplitz matrix~\cite{deifts2011asymptotics}, with symbol
$\Theta_{\mathrm H}(k_F-|k|)$. The second term in~\eqref{eq:gs-corr} depends
only on $n+m$, which defines a so-called Hankel matrix. 
Thus, the full correlator exhibits a Toeplitz-plus-Hankel structure. 

Crucially, for $|h|> 2$ the symbol of the correlator in~\eqref{eq:gs-corr} 
is smooth as a function of $k$. On the other hand, for $|h|<2$ it exhibits
a jump discontinuity at $\pm k_F$, which is the main signature of
critical behavior. In this case we have a logarithmic violation of the area law
in the ground-state entanglement 
entropies~\cite{calabrese2009entanglemententropy,laflorencie2016quantum}. 
In the following sections, by using the approach of
Ref.~\cite{dabbruzzo2021self} we will show that in the presence of thermal baths 
locally coupled to the chain, the steady-state fermionic correlator  
exhibits a similar structure as in Eq.~\eqref{eq:gs-corr}. In particular, 
even though the symbol of the correlator is affected by the presence of the 
bath, it is not smooth as a function of $k$. This gives rise to 
logarithmic scaling of the steady-state mutual information.


\subsubsection{Kitaev chain} \label{sec:kitaev}

Let us now consider the Kitaev chain~\cite{Kitaev_2001}.
This is obtained from~\eqref{eq:ham-gen} by choosing 
\begin{subequations}
	\begin{align}
		Q_{nm} &= -J\delta_{n,m-1}-J\delta_{n,m+1}-h\delta_{n,m}, \\ 
		P_{nm} &= -\Delta\delta_{n,m-1}+\Delta\delta_{n,m+1},
	\end{align}
\end{subequations}
with $J$ the hopping strength, $h$ a magnetic field, and $\Delta$ the strength
of the pairing term. In the following, we will set $\Delta=J=1$. 
Thus, the Hamiltonian of the Kitaev chain reads as 
\begin{equation} \label{eq:kit-ham}
	H_\mathcal{S} = -\sum_{n=1}^N \left( a_n^\dagger a_{n+1}
	+ a_n^\dagger a_{n+1}^\dagger + \text{h.c.} \right)
	- h \sum_{n=1}^N a_n^\dagger a_n.
\end{equation}
For this model we exclusively employ PBC, choosing $a_{N+1} = a_1$.
The Hamiltonian~\eqref{eq:kit-ham} can be rewritten as in~\eqref{eq:ek} 
with single-particle dispersion 
\begin{equation} \label{eq:kit-disp}
	\omega_k = \sqrt{h^2 + 4h\cos(k) + 4}, 
\end{equation}
where the index $k$ is chosen as in~\eqref{eq:k_bc} for PBC.
The functions $\phi_{jk}$ and $\psi_{jk}$ [cf.~Eqs.~\eqref{eq:phipsi-def}]
that encode the Fourier transform and the Bogoliubov transformation 
needed to diagonalize~\eqref{eq:kit-ham} are given by
\begin{subequations} \label{eq:phipsi-is}
	\begin{align}
		\label{eq:phi-is} \phi_{nk} &= \frac{e^{ikn}}{\sqrt{N}}, \\
		\label{eq:psi-is} \psi_{nk} &= e^{i\xi(k)}\phi_{nk},
	\end{align}
\end{subequations}
where we defined 
\begin{equation} \label{eq:xi-def}
	e^{i\xi(k)} \coloneqq -\frac{h+2e^{ik}}{\omega_k}
\end{equation}
and the so-called Bogoliubov angle $\xi(k) \in \mathbb{R}$ is
\begin{subequations} \label{eq:xi}
	\begin{align}
		\label{eq:cosxi}
		\cos\xi(k) &= -\frac{h+2\cos k}{\sqrt{h^2 + 4h\cos k + 4}}, \\
		\label{eq:sinxi}
		\sin\xi(k) &= \frac{2\sin k}{\sqrt{h^2 + 4h\cos k + 4}}.
	\end{align}
\end{subequations}
It is clear that $\xi(k)$ is continuous as a function of $k$, for $|h| \neq 2$.
For $h=2$, a jump discontinuity appears at $k=\pm\pi$, while for $h=-2$
it emerges at $k=0$:
this is the transition between trivial $(|h| > 2)$
and topological phase $(|h| < 2)$.
As for the tight-binding chain (see Sec.~\ref{sec:tb}), 
at $|h|=2$ long-wavelength properties of the 
ground state of the Kitaev chain are described by a CFT with central 
charge $c=1/2$. Consequently, the ground state exhibits logarithmic 
violations of the area law for the entanglement entropy. 
Again, below we show that the singular structure of Eqs.~\eqref{eq:xi} 
survives in the presence of localized baths, 
giving rise to logarithmic scaling of the mutual information.


\subsubsection{Majorana correlation function} \label{sec:majo}

To determine entanglement-related quantities, it is convenient to introduce the 
Majorana operators~\cite{VidalLatorre_2003, VidalLatorre_2004}: 
\begin{equation} \label{eq:majo-def}
	w_{2n-1} = \frac{1}{\sqrt{2}}\left( a_n^\dagger + a_n \right),
	\quad
	w_{2n} = \frac{i}{\sqrt{2}} \left( a_n^\dagger - a_n \right).
\end{equation}
It is straightforward to write these in terms of the Bogoliubov operators $b_k$
that diagonalize the models [cf.~\eqref{eq:ek}]:
\begin{equation}
	\begin{bmatrix}
		w_{2n-1} \\ w_{2n}
	\end{bmatrix} = \frac{1}{\sqrt{2}} \sum_k \begin{bmatrix}
		\phi_{nk} & \phi^*_{nk} \\ -i\psi_{nk} & i\psi^*_{nk}
	\end{bmatrix} \begin{bmatrix}
		b_k \\ b_k^\dagger
	\end{bmatrix},
\end{equation}
where $\phi_{nk}$ and $\psi_{nk}$ are given in Eq.~\eqref{eq:phipsi-def}. 
For the tight-binding chain and the Kitaev chain $\phi_{nk},\psi_{nk}$ are
reported in Eqs.~\eqref{eq:phipsi-xx} and~\eqref{eq:phipsi-is}, respectively. 
One can show that the generic expectation value $\langle w_a w_b\rangle$ is 
written as
\begin{equation}
	\langle w_a w_b \rangle = \frac{\delta_{ab} + i\Gamma_{ab}}{2},
\end{equation}
where $\Gamma$ is a $2N \times 2N$ matrix of the form
\begin{equation} \label{eq:gamma-block-struct}
	\Gamma = \begin{bmatrix}
		\Pi_{11} & \Pi_{12} & \cdots & \Pi_{1N} \\
		\Pi_{21} & \Pi_{22} & \cdots & \Pi_{2N} \\
		\vdots & \vdots & \ddots & \vdots \\
		\Pi_{N1} & \Pi_{N2} & \cdots & \Pi_{NN}
	\end{bmatrix},
\end{equation}
with $\Pi_{nm}$ being a $2 \times 2$ block defined by
\begin{equation} \label{eq:gamma-def}
	\Pi_{nm} \coloneqq \begin{bmatrix}
		0 & \Re[\phi\,\theta\,\psi^\dagger]_{nm} \\
		-\Re[\phi\,\theta\,\psi^\dagger]_{mn} & 0
	\end{bmatrix}.
\end{equation}
In writing~\eqref{eq:gamma-def} we assumed the
matrix $K_{kq} \coloneqq \langle b_k^\dagger b_q \rangle$ to be
diagonal and the matrix $F_{kq} \coloneqq \langle b_k b_q \rangle$
to be zero, which will turn out to be true in our formalism
(see Sec.~\ref{sec:lindblad}). 
Here $\theta_{kq}$ is the occupation of the Bogoliubov modes $b_k$ given by 
\begin{equation} \label{eq:theta-def}
	\theta_{kq} = \delta_{kq} \big( 1-2\langle b^\dagger_k b_k\rangle \big).
\end{equation}
Notice that $\Gamma$ is a real skew-symmetric matrix 
of even dimension. This means that it has pairs of 
eigenvalues $\pm i \nu_r$ with $\nu_r \in \mathbb{R}$.

Let us now specialize the matrix $\Gamma$ to the case of 
the tight-binding chain (see section~\ref{sec:tb}). 
By using Eqs.~\eqref{eq:phipsi-xx} in~\eqref{eq:gamma-def} we obtain
\begin{equation} \label{eq:G-def}
	G_{nm} \coloneqq \Re [\phi \, \theta \, \psi^\dagger]_{nm} 
	= \delta_{nm} - 2 \, C_{nm},
\end{equation}
where $C_{nm}=\langle a^\dagger_n a_m\rangle$ is the fermion correlation 
function. This implies that 
\begin{equation} \label{eq:GC}
	\Gamma = G \otimes \begin{bmatrix}
    0 & 1 \\ -1 & 0
  \end{bmatrix},
\end{equation}
from which we conclude that the eigenvalues of 
$\Gamma$ are $\pm i \nu_r$ if and only if $\nu_r$ are the eigenvalues of $G$.

Let us now consider the Kitaev chain with PBC.
By using Eqs.~\eqref{eq:phipsi-is} we obtain 
\begin{equation} \label{eq:re-kit}
	\Re[\phi \,\theta \,\psi^\dagger]_{nm} = \frac{1}{N} \Re \sum_k
	\theta_{kk} e^{-i\xi(k)} e^{i k(n-m)}, 
\end{equation}
where $\xi(k)$ is defined in~\eqref{eq:xi-def}, and $\theta_{kk}$ is 
given in~\eqref{eq:theta-def}.
Using the fact that $\theta_{kk}$ is an even function of $k$ and  
$\xi(k)$ is an odd one, we can write  
\begin{equation} \label{eq:gamma-is}
	\Pi_{nm} = \int_{-\pi}^\pi \frac{dk}{2\pi} 
	\begin{bmatrix}
		0 & \theta_{kk} e^{-i\xi(k)} \\ -\theta_{kk}e^{i\xi(k)} & 0
	\end{bmatrix}
  	e^{i k(n-m)}, 
\end{equation}
where we took the thermodynamic limit $N \to \infty$. 
Eq.~\eqref{eq:gamma-is} holds for a generic thermodynamic 
state, which is characterized by the functions $\theta_{kk}$. 
Like for the tight-binding chain, the ground-state of the Kitaev chain is
the state annihilated by all the Bogoliubov operators $b_k$.
In this case $\theta_{kk}=1$ [cf.~\eqref{eq:theta-def}],  
and the ground-state is characterized by
\begin{equation} \label{eq:gamma-gs}
	\Pi^{(\mathrm{GS})}_{nm} = \int_{-\pi}^\pi \frac{dk}{2\pi} 
	\begin{bmatrix}
		0 & e^{-i\xi(k)} \\ -e^{i\xi(k)} & 0
	\end{bmatrix}
 	e^{i k(n-m)}.
\end{equation}
In the presence of external thermal baths, the Majorana correlator $\Gamma$ is 
determined by~\eqref{eq:gamma-is}, with $\theta_{kk}$ encoding
the properties of the baths.


\subsection{Entropy in free-fermion systems} \label{sec:ent-free-f}

For free-fermion systems, the von Neumann entropy of a subsystem
$A$ of length $\ell$ (see Fig.~\ref{fig:sb}), and the R\'enyi entropies in 
general~\cite{peschel2009reduced}, are obtained from the Majorana
correlation matrix $\Gamma_A$, which is obtained 
from~\eqref{eq:gamma-block-struct} by restricting $n,m \in A$.
If $\pm i \nu_r$ are the eigenvalues of $\Gamma_A$, then
\begin{equation} \label{eq:S-nu}
	S_A = \sum_{r=1}^\ell e(1,\nu_r), 
\end{equation}
where we defined the function $e(x,\nu)$ as 
\begin{equation} \label{eq:e-def}
	e(x,\nu) \coloneqq -\frac{x-\nu}{2}\ln(\frac{x-\nu}{2}) - 
	\frac{x+\nu}{2}\ln(\frac{x+\nu}{2}).
\end{equation}
Notice that Eq.~\eqref{eq:S-nu} is well-defined because
one can show that $-1 \leq \nu_r \leq 1$.

\begin{figure}
	\centering
	\includegraphics[scale=0.75]{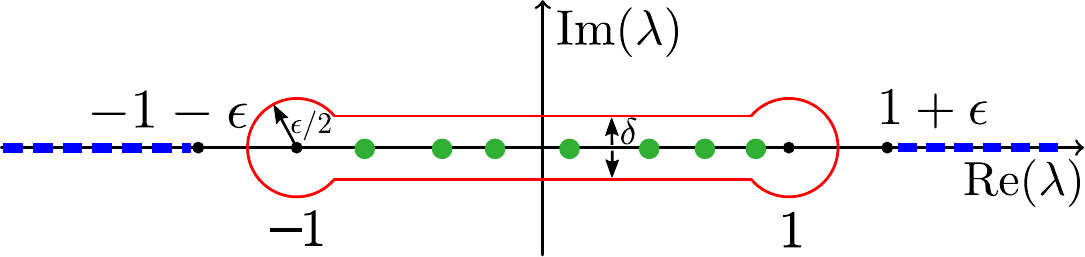}
	\caption{
		Contour $\gamma$ in the complex plane for $\lambda$ used to compute
		the von Neumann entropy of an interval [cf.~Eq.~\eqref{eq:ent-C}].
		Dashed lines at $(-\infty,-1-\epsilon]\cup [1+\epsilon,\infty)$ denote
		a branch cut. Dots in the region $[-1,1]$ are the zeros of
		$D_\ell(\lambda)$ [cf.~Eq.~\eqref{eq:D-def}], 
		or, equivalently, the poles of $d D_\ell(\lambda)/d\lambda$.
		Here we are interested in the limits $\epsilon\to 0$ and $\delta \to 0$.
	}
	\label{fig:contour}
\end{figure}

It is convenient to rewrite the sum in Eq.~\eqref{eq:S-nu} as 
an integral in the complex plane. To this purpose we define the determinant 
\begin{equation} \label{eq:D-def}
	D_\ell(\lambda) \coloneqq
	\det\left(\lambda\mathds{1}-i\Gamma_A\right).
\end{equation}
A straightforward application of Cauchy's theorem allows to 
rewrite Eq.~\eqref{eq:S-nu} as
\begin{equation} \label{eq:ent-C}
	S_A = \lim_{\delta, \epsilon \rightarrow 0^+} \frac{1}{4\pi i}
	\oint_\gamma d\lambda \, e(1+\epsilon,\lambda)
	\frac{d \ln D_\ell(\lambda)}{d\lambda},
\end{equation}
where we used the fact that $e(1,\nu)=e(1,-\nu)$.
The contour $\gamma$ in the complex plane is shown in Fig.~\ref{fig:contour}. 
Dashed blue lines in the figure are the branch cuts of $e(1+\epsilon,\lambda)$ 
at $(-\infty,-1-\epsilon]\cup [1+\epsilon,\infty)$.
The horizontals parts of the contour are shifted by $\delta$ from the real axis. 
Finally, the function $d\ln(D_\ell)/d\lambda$ 
has simple poles in the interval $[-1,1]$ (green dots in the figure). 
In the limit $\ell\to\infty$ the poles become dense, forming a new branch cut. 
The strategy~\cite{jin2004quantum} to obtain the asymptotic scaling
of $S_A$ in the limit $\ell=|A|\to\infty$ 
is to first obtain $D_\ell$ in the limit $\ell\to\infty$,
then using it in Eq.~\eqref{eq:ent-C}.

\section{Global Lindblad master equation} \label{sec:lindblad}

To make the paper self contained, we now recap the formalism used to treat
self-consistently thermal baths in the Lindblad approximation,
within quadratic models~\cite{dabbruzzo2021self}. 
Let us consider the interaction between the fermionic chain
$\mathcal{S}$ and the environment $\mathcal{E}$ (see Fig.~\ref{fig:sb}).
The global system $\mathcal{U}={\mathcal S}\cup \mathcal{E}$
is described by the Hamiltonian
$H_\mathcal{U} = H_\mathcal{S} \otimes \mathds{1}_\mathcal{E} +
\mathds{1}_\mathcal{S} \otimes H_\mathcal{E} + H_I$,
where $H_\mathcal{E}$ is the Hamiltonian of the environment and
$H_I$ models the interaction between system and environment.
We can always write $H_I$ in the form
\begin{equation} \label{eq:se-ham}
	H_I = \sum_\alpha O_\alpha \otimes R_\alpha,
\end{equation}
where $O_\alpha$ and $R_\alpha$ are Hermitian operators 
acting on $\mathcal{S}$ and $\mathcal{E}$, respectively. 
In the following we restrict ourselves to the situation in which 
$O_\alpha$ act nontrivially only on a finite number of sites of the 
chain (see Fig.~\ref{fig:sb}). 

Given a state $\rho(t)$ of the entire system $\mathcal{S}\cup \mathcal{E}$, 
we are interested in the evolution of the reduced density matrix
$\rho_\mathcal{S}(t) \coloneqq \Tr_\mathcal{E}[\rho(t)]$.
In the Markovian regime, the dynamics is described by a Lindblad master 
equation of the form~\cite{petruccione2002the} 
\begin{equation} \label{eq:lin-eq}
	\frac{d\rho_\mathcal{S}(t)}{dt} = -i[H, \rho_\mathcal{S}(t)] 
	+ \mathcal{D}(\rho_\mathcal{S}(t)),
\end{equation}
where both $H$ and $\mathcal{D}$ have  to be determined. Let us assume
that the environment consists of a finite number of uncorrelated
fermionic infinite thermal baths, such that 
\begin{equation}
	H_\mathcal{E} = \sum_\alpha \int dp \, \varepsilon_{\alpha,p} \,
	d^\dagger_{\alpha,p} d_{\alpha,p},
\end{equation}
where $\alpha$ is here an index that labels the bath, $d_{\alpha,p}$ the 
fermionic operators of the bath, and $\varepsilon_{\alpha,p}$ the 
bath dispersion. We also consider a generic linear coupling 
[cf.~Eq.~\eqref{eq:se-ham}] between the system and the baths:
\begin{align}
	\label{eq:o}
	O_\alpha &= \sum_{j \in \mathcal{I}_\alpha}
	\left( a_j + a_j^\dagger \right),\\
	\label{eq:r}
	R_\alpha &= \int dp \, g_{\alpha,p} \left( d_{\alpha,p} +
	d^\dagger_{\alpha,p} \right),
\end{align}
where $g_{\alpha,p}$ is the strength of the coupling and $\mathcal{I}_\alpha$
are the sites of $\mathcal{S}$ that are coupled to the bath $\alpha$. 
Here we focus on the situation in which each bath is coupled to a single 
site of the system. 
For instance, for the case in Fig.~\ref{fig:sb}(a), one has $\alpha=1,2$ 
with $\mathcal{I}_1=\{1\}$ and $\mathcal{I}_2=\{N\}$. 

The dissipator $\mathcal{D}$ in Eq.~\eqref{eq:lin-eq} can be written
as~\cite{dabbruzzo2021self}
\begin{multline} \label{eq:dissipator1}
	\mathcal{D}(\rho) = \sum_{\alpha,k} \Phi_{\alpha,k}
	\Big[ \Gamma_\alpha(\omega_k) \left( 2b_k \rho b_k^\dagger -
	\{b_k^\dagger b_k, \rho\} \right) \\
	+ \Gamma_\alpha(-\omega_k) \left( 2b_k^\dagger \rho b_k -
	\{b_k b_k^\dagger, \rho\} \right) \Big].
\end{multline}
For simplicity we removed the subscript $\mathcal{S}$ in $\rho_\mathcal{S}$.
The $b_k$'s denote the Bogoliubov operators that diagonalize the system
[cf.~Eq.~\eqref{eq:bk-def}], whereas $\omega_k$ are the corresponding single-particle
energies [cf.~Eq.~\eqref{eq:ek}]. Even though the interaction Hamiltonian
$H_I$ is local in space, the dissipator $\mathcal{D}(\rho)$ is written in terms
of nonlocal operators. In constrast, with common approaches, the Lindblad
operators are chosen \emph{ad hoc} and are typically local. Information about
locality of the baths is encoded in the functions
\begin{equation}
	\Phi_{\alpha,k} \coloneqq \bigg|
	\sum_{j \in \mathcal{I}_\alpha} \phi_{jk} \bigg|^2.
\end{equation}
Moreover, we have defined~\cite{petruccione2002the}
\begin{equation}
	\Gamma_\alpha(\omega) = \begin{cases}
		J_\alpha(\omega)(1-f_\alpha(\omega)) & \omega > 0, \\
		J_\alpha(-\omega)f_\alpha(-\omega) & \omega < 0,
	\end{cases}
\end{equation}
which is written in terms of the Fermi-Dirac distribution $f_\alpha$ associated
with the bath $\alpha$ at temperature $T_\alpha$ and
chemical potential $\mu_\alpha$,
\begin{equation} \label{eq:fd}
	f_\alpha(\omega) = \frac{1}{1 + e^{\beta_\alpha(\omega-\mu_\alpha)}},
	\quad
	\beta_\alpha = \frac{1}{T_\alpha},
\end{equation}
and the spectral density of the bath $\alpha$,
\begin{equation} \label{eq:J-alpha}
	J_\alpha(\omega) = \pi \int dp \, |g_{\alpha,p}|^2
	\delta(\omega - \varepsilon_{\alpha,p}).
\end{equation}
In the chosen diagonalization scheme we always have $\omega_k \geq 0$,
hence we can also rewrite
\begin{multline} \label{eq:D-self}
	\mathcal{D}(\rho) = \sum_{\alpha,k} \gamma_{\alpha,k} \Big[
	(1-f_\alpha(\omega_k))
	\left( 2b_k \rho b_k^\dagger - \{b_k^\dagger b_k, \rho\} \right) \\
	+ f_\alpha(\omega_k) \left( 2b_k^\dagger \rho b_k -
	\{b_k b_k^\dagger, \rho\} \right) \Big],
\end{multline}
where
\begin{equation} \label{eq:gamma-bath}
	\gamma_{\alpha,k} \coloneqq J_\alpha(\omega_k) \Phi_{\alpha,k}.
\end{equation}

Besides the dissipative effect encoded in $\mathcal{D}(\rho)$, the presence 
of the baths also renormalizes the unitary part of the
Lindblad equation~\eqref{eq:lin-eq}. 
Indeed, the effective Hamiltonian $H$ reads 
\begin{equation}
	H = \sum_k \widetilde{\omega}_k b_k^\dagger b_k, 
\end{equation}
where the ``dressed'' single-particle dispersion $\widetilde\omega_k$ reads
\begin{equation} \label{eq:tildeomega}
	\widetilde{\omega}_k = \omega_k \Bigg( 1+ \frac{2}{\pi}\sum_\alpha
	\Phi_{\alpha,k} 
	\mathcal{P}\!\int_0^\infty \!\!\!d\epsilon
  \frac{J_\alpha(\epsilon)}{\omega_k^2 - \epsilon^2} \Bigg),
\end{equation}
with $\mathcal{P}$ denoting Cauchy's principal value.

Crucially, the Lindblad equation~\eqref{eq:lin-eq} is derived by using a
\emph{full secular approximation}~\cite{petruccione2002the,dabbruzzo2021self,
dabbruzzo2021self2}, which neglects rapidly oscillating terms
$\propto \exp(i (\omega_k-\omega_{k'})t)$.
Moreover, we neglect degeneracy in the spectrum,
assuming that $\omega_k \neq \omega_{k'}$ if $k \neq k'$.
Both these approximations are in general uncontrolled, and checking
their validity would require an \emph{ab initio} treatment of the baths. 

The master equation~\eqref{eq:lin-eq} is quadratic in the Bogoliubov operators
$b_k, b_k^\dagger$. This means that if the state of the system is 
Gaussian at a certain initial time, it will remain Gaussian 
at all subsequent times. Therefore, the state $\rho$ is completely 
determined by the two-point functions of the Majorana
fermions~\eqref{eq:majo-def}. Equivalently, one can use the correlators
$K_{kq}$ and $F_{kq}$ defined as 
\begin{subequations}
	\begin{align}
		K_{kq} & \coloneqq \mathrm{Tr}[\rho \, b^\dagger_k b_q], \\
  		F_{kq} & \coloneqq \mathrm{Tr}[\rho \, b_k b_q].
	\end{align}
\end{subequations}
A direct computation allows to obtain the evolution of $K_{kq}$ 
and $F_{kq}$ as~\cite{dabbruzzo2021self}  
\begin{subequations}
	\begin{multline} \label{eq:Ckq}
		\frac{dK_{kq}}{dt} = \Big[ i(\widetilde{\omega}_k -
		\widetilde{\omega}_q) - \sum_\alpha (\gamma_{\alpha,k} +
		\gamma_{\alpha,q}) \Big] K_{kq}(t) \\
		+ 2\delta_{kq} \sum_\alpha \gamma_{\alpha,k} f_\alpha(\omega_k),
	\end{multline}
	\begin{equation} \label{eq:Fkq}
		\frac{dF_{kq}}{dt} = \Big[ -i(\widetilde{\omega}_k +
		\widetilde{\omega}_q) - \sum_\alpha (\gamma_{\alpha,k} +
		\gamma_{\alpha,q}) \Big] F_{kq}(t).
	\end{equation}
\end{subequations}
Here $\widetilde\omega_k$ are the modified single-particle energies in
Eq.~\eqref{eq:tildeomega}, the rates $\gamma_{\alpha,k}$ are defined in
Eq.~\eqref{eq:gamma-bath}, and $f_{\alpha}(\omega)$ is the Fermi-Dirac
distribution of the bath [cf.~Eq.~\eqref{eq:fd}]. 
We anticipate that, for the setting in Fig.~\ref{fig:sb},
due to the simple structure of Eqs.~\eqref{eq:phipsi-xx}
and~\eqref{eq:phipsi-is}, the dependence on $\gamma_{\alpha,k}$ drops out.
Assuming that $\gamma_{\alpha,k}$ are not all equal to zero (which is
obviously true if the system is actually coupled to the environment),
in the stationary limit $t \rightarrow \infty$ we obtain
\begin{equation} \label{eq:statio}
	K_{kq} = \delta_{kq}\frac{\sum_\alpha \gamma_{\alpha,k} f_\alpha(\omega_k)}
	{\sum_\alpha \gamma_{\alpha,k}}, \quad F_{kq}=0.
\end{equation}
Thus the correlation function in momentum space becomes diagonal, and it is 
a convex combination of the Fermi-Dirac distributions of the baths. 
Equation~\eqref{eq:statio} is the main ingredient to extract steady-state 
properties of the system (see Sec.~\ref{sec:tb-ent} and
Sec.~\ref{sec:kitaev-ent}). Note that, if the baths are identical
($f_\alpha$ does not depend on $\alpha$), we obtain
$K_{kq}=\delta_{kq}f(\omega_k)$. 
Interestingly, even in this situation, the statistical ensemble that describes 
the steady state is not the standard finite-temperature ensemble of the
underlying free-fermion model, due to the nonzero 
chemical potential in Eq.~\eqref{eq:fd}. As we will show in the following,
this implies that the steady-state von Neumann entropy exhibits logarithmic
additive corrections to the expected volume-law scaling at finite temperature.


\section{Scaling of entropy in the tight-binding chain} \label{sec:tb-ent}

In this section we derive the scaling equation~\eqref{eq:ent-asy}
of the steady-state von Neumann entropy for a subinterval of the
tight-binding chain [cf.~Eq.~\eqref{eq:ham-xx}].
A similar calculation was performed in Refs.~\cite{Eisler2014,Kormos2017},
but for a nonequilibrium setting which is different from ours.

First of all we recall that, from Eq.~\eqref{eq:GC}, $\Gamma$ has
eigenvalues $\pm i\nu_r$ if and only if $G$ [cf.~Eq.~\eqref{eq:G-def}]
has eigenvalues $\nu_r$. We can exploit this fact by expressing the contour
integral~\eqref{eq:ent-C} in terms of
\begin{equation}
	B_\ell(\lambda) \coloneqq \det(\lambda \mathds{1} - G_A),
\end{equation}
finding
\begin{equation} \label{eq:new-contour}
	S_A = \lim_{\delta,\epsilon \to 0^+} \frac{1}{2\pi i} \oint_\gamma
	d\lambda \, e(1+\epsilon,\lambda) \frac{d \ln B_\ell(\lambda)}{d\lambda},
\end{equation}
where $G_A$ is the matrix obtained from $G$ by restricting indices to
$n,m \in A$, and $\ell = |A|$.
Using the definitions of $\phi_{nk}$ and $\psi_{nk}$ reported 
in Eqs.~\eqref{eq:phipsi-xx}, we find, in the limit $N \to \infty$,
\begin{equation} \label{eq:gnm}
	G_{nm} = \int_{-\pi}^\pi \frac{dk}{2\pi} \,
	\widetilde\theta_{kk} \left[ e^{ik(n-m)} - \zeta e^{ik(n+m)} \right],
\end{equation}
where $\zeta=0$ and $\zeta=1$ corresponds to PBC and OBC, respectively. 
This equation defines a Toeplitz matrix for $\zeta=0$, whereas one 
has a Toeplitz-plus-Hankel matrix~\cite{deifts2011asymptotics} for $\zeta=1$. 
The so-called symbol of $G_{nm}$ is
\begin{equation} \label{eq:tk}
	\widetilde\theta_{kk} \coloneqq
	\theta_{kk}\,\mathrm{sgn}( -h-2\cos k ), 
\end{equation}
where $\theta_{kk}$ is given in Eq.~\eqref{eq:theta-def} and the sign function 
is the same as in Eq.~\eqref{eq:psi-xx}. The function $\theta_{kk}$ encodes 
the information about the steady state and is obtained from~\eqref{eq:statio}. 
The function $\widetilde\theta_{kk}$ is smooth, except for 
the sign function which displays a singularity for $|h|<2$,
thus giving rise to logarithmic corrections to the von Neumann entropy.
Since the sign function comes from the Lieb-Schultz-Mattis
matrices~\eqref{eq:phipsi-xx}, the attentive reader could think that its
appearance in the Majorana correlation matrix is an artifact originated from
our choice of using the Bogoliubov modes to 
diagonalize the tight-binding Hamiltonian. However, in App.~\ref{app:fourier}
we show that, if one constructs the master equation in terms of Fourier modes
(which still diagonalize the Hamiltonian, but with continuous coefficients),
then the discontinuity arises from their steady-state correlation function.
As expected, physical quantities are not affected by this change and
Eqs.~\eqref{eq:gnm} and~\eqref{eq:tk} are unaltered.

To extract the scaling behavior of the mutual information~\eqref{eq:mi-intro}
one has to determine the asymptotic scaling of the von Neumann entropy 
$S_A$ for a subsystem $A$ of length $\ell\to\infty$. To that purpose,
we first study the asymptotic behavior of 
$B_\ell(\lambda) \equiv D_\ell[g_\lambda]$,
being the determinant of the Toeplitz (or Toeplitz-plus-Hankel) matrix with
symbol $g_\lambda(k) \coloneqq \lambda-\widetilde\theta_{kk}$ given by 
\begin{equation} \label{eq:tilde-simb}
	g_\lambda(k)=\left\{
	\begin{array}{cl}
	\lambda-\theta_{kk} & \;\; k\in[-\pi,-k_F]\cup [k_F,\pi) ,\\
		\lambda+\theta_{kk} & \;\; k\in[-k_F,k_F) .
	\end{array}
	\right.
\end{equation}
Since this symbol may have jump discontinuities at $\pm k_F$, we use the
Fisher-Hartwig theorem to extract the scaling of $D_\ell[g_\lambda]$
for $\ell \to \infty$. The result is then substituted back in
Eq.~\eqref{eq:new-contour}, from which the expression of the von Neumann
entropy emerges. The details of the computation are reported in
App.~\ref{app:tight-binding}, where we find
\begin{equation} \label{eq:ent-expansion}
	S_A = \alpha \ell + \frac{c(\Theta)}{3\nu} \ln(\ell) + \mathcal{O}(1),
\end{equation}
where $\nu = 1$ for PBC and $\nu = 2$ for OBC.
The prefactor of the linear term,
\begin{equation} \label{eq:lin-coeff}
	\alpha = -\int_{-\pi}^\pi \frac{dk}{2\pi}
	\Big[ K_{kk}\ln(K_{kk}) + (1-K_{kk}) \ln(1-K_{kk}) \Big],
\end{equation}
is the von Neumann entropy per volume of the full system in the limit
$N \to \infty$, i.e.,
\begin{equation}
	\alpha = \lim_{\ell \to \infty} \frac{S_A}{\ell} =
	\lim_{N \to \infty} \frac{S_N}{N},
\end{equation}
where we denoted with $S_N$ the entropy of the full system.
For pure states one has either $K_{kk} = 0$ or $K_{kk} = 1$, which implies
that $S_N = 0$ and $\alpha = 0$, as it should be. This is not the case
in the presence of the environment, because the state is not pure.
A similar behavior is typically observed in generic out-of-equilibrium 
quadratic fermionic and bosonic systems~\cite{alba2021spreading,
carollo2022dissipative,alba2022hydrodynamics,alba2022logarithmic}.

\begin{figure}
	\centering
	\includegraphics[scale=0.8]{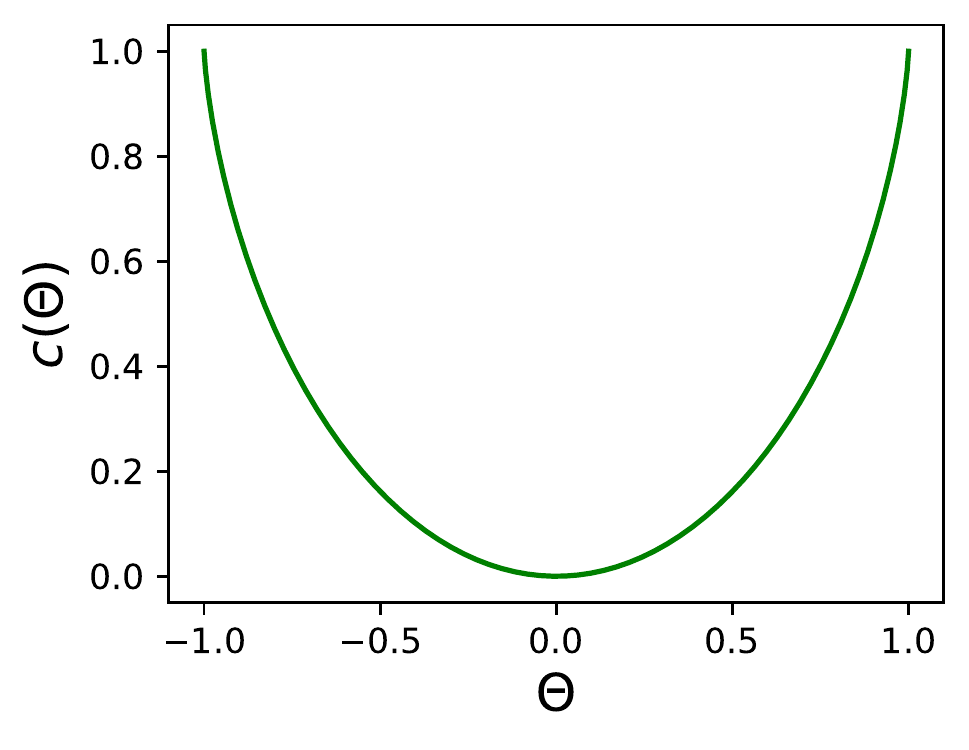}
	\caption{
		Effective central charge $c(\Theta)$ in Eq.~\eqref{eq:ctheta-int}
		for the tight-binding chain coupled to localized thermal baths.
		Here $\Theta \in [-1,1]$ encodes the information about the baths.
		For a single bath at the left edge of the chain, $\Theta$ is given by
		Eq.~\eqref{eq:theta-single}. 
		For two baths at the edges of the chain, $\Theta$ is given by
		Eq.~\eqref{eq:theta-is-3}. 
		In the limit $|\Theta|\to1$, one recovers the CFT result $c=1$. 
		For $\Theta \to 0$ one has $c \to 0$, which corresponds to a
		high-temperature limit.
	}
	\label{fig:theo}
\end{figure}

The prefactor $c(\Theta)$ of the logarithm is nonvanishing only when the
symbol~\eqref{eq:tilde-simb} has the jump discontinuity
(which happens precisely when the model is critical), and it is given by
\begin{equation} \label{eq:ctheta-int}
	c(\Theta) = \frac{3}{\pi^2}\! \left[ (1+\Theta)
	\mathrm{Li}_2\!\left( \frac{2\Theta}{\Theta+1} \right) \!+\!
	(1-\Theta) \mathrm{Li}_2\!\left( \frac{2\Theta}{\Theta-1}
	\right) \right] \!,
\end{equation}
where $\Theta \coloneqq \theta_{kk}\eval_{k=k_F}$ is the steady-state
density of Bogoliubov excitations at the Fermi level, which contains
information about the environment, and
\begin{equation} \label{eq:dilog}
	\mathrm{Li}_2(x) = -\int_0^x dz \, \frac{\ln(1-z)}{z}
\end{equation}
is the dilogarithm function~\cite{nist2022digital}. This result is
somewhat reminiscent of the effective central charge obtained in
free-fermion chains in the presence of 
defects~\cite{eisler2010entanglement,eisler2012on,calabrese2011entanglement,
calabrese2011the,calabrese2012entanglement}.
Formally, this expression is a special symmetric case of the result
obtained in Ref.~\cite{Eisler2014}, which is valid for two arbitrary
asymmetric jumps in the symbol.
For $\Theta \to 1$, which corresponds to the ground state of the
closed chain, one recovers the standard central charge $c=1$, since
$\mathrm{Li}_2(1) = \pi^2/6$. Instead, in the opposite limit $\Theta \to 0$
one obtains that $c(\Theta)$ vanishes, since $\mathrm{Li}_2(0) = 0$.
The behavior of $c(\Theta)$ as a function of $\Theta$ is illustrated
in Fig.~\ref{fig:theo}.

\subsection{Single bath} \label{sec:xx-single}

To illustrate our results, we first focus on the tight-binding chain with OBC
and with only one edge coupled to a bath with temperature $T_L$ and
chemical potential $\mu_L$ (see Fig.~\ref{fig:sb}).
In the steady state, from Eq.~\eqref{eq:statio} we have
\begin{equation} \label{eq:prev}
	K_{kq} = \langle b^\dagger_k b_q\rangle=\delta_{kq}f(\omega_k),
\end{equation}
where $\omega_k$ are the single-particle energies
[cf.~Eq.~\eqref{eq:xx-obc-disp}], and $f(\omega_k)$ is the Fermi-Dirac
distribution describing the bath [cf.~Eq.~\eqref{eq:fd}].  
Eq.~\eqref{eq:prev} implies that the function $\theta_{kk}$
[cf.~\eqref{eq:theta-def}] is given by
\begin{equation}
	\theta_{kk} = 1-2f(\omega_k). 
\end{equation}
Thus, we have 
\begin{equation} \label{eq:theta-single}
	\Theta = \theta_{kk}\eval_{k=k_F} = 1-2f(0)
	= \tanh(-\frac{\mu_L}{2T_L}).
\end{equation}
Notice that $\Theta$ depends only on the ratio $\mu_L/T_L$. 
The limit $\mu_L/T_L\to 0$ gives $\Theta \to 0$ and a vanishing 
$c(\Theta)$ (see Fig.~\ref{fig:theo}), so that the logarithmic correction
to the entropy disappears.
This regime corresponds to either infinite temperature $T_L \to \infty$
or vanishing chemical potential $\mu_L \to 0$.
On the other hand, for $|\mu_L/T_L| \to \infty$ we have $|\Theta| \to 1$, so 
that $c \to 1$. Moreover, in this limit $K_{kk} \to 0$ or $K_{kk} \to 1$, which 
implies that $\alpha\to0$ [cf.~\eqref{eq:lin-coeff}]. Thus, 
we recover the ground-state scaling of the von Neumann entropy.


\subsection{Two baths} \label{sec:xx-two-baths}

Let us now discuss the case with two different thermal baths, first considering
the simpler case of PBC. Under the assumption that the spectral density
$J(\omega)$ [cf.~\eqref{eq:J-alpha}] is the same for the two baths,
we have for the couplings
\begin{equation}
	\gamma_{L,k} = \gamma_{R,k} = \frac{J(\omega_k)}{N},
\end{equation}
since, from Eq.~\eqref{eq:phi-xx}, we have
$|\phi_{1k}|^2 = |\phi_{Nk}|^2 = 1/N$. Using Eq.~\eqref{eq:statio},
we therefore obtain
\begin{align}
	\label{eq:Ckq-1}
	K_{kq} &= \delta_{kq}\frac{f_L(\omega_k) + f_R(\omega_k)}{2},\\
	\label{eq:theta-is-2}
	\theta_{kk} &= \frac{1}{2} \left[
	\tanh\left(\frac{\omega_k-\mu_L}{2T_L}\right)
	+\tanh\left(\frac{\omega_k-\mu_R}{2T_R}\right) \right].
\end{align}
Notice that the correlator $K_{kq}$ does not depend on the 
couplings $\gamma_{L/R,k}$. Moreover, as is clear from~\eqref{eq:Ckq-1}, 
the steady-state correlator  is written in terms of the average between the 
Fermi-Dirac distributions describing the baths. 
From Eq.~\eqref{eq:theta-is-2} we obtain 
\begin{equation} \label{eq:theta-is-3}
	\Theta = \frac{1}{2}\left[\tanh\left(-\frac{\mu_L}{2T_L}\right)
	+\tanh\left(-\frac{\mu_R}{2T_R}\right)\right].
\end{equation}
Notice that $\Theta$ depends only the ratios $\mu_L/T_L$ and $\mu_R/T_R$,
which is structurally similar to what we obtained in the
single-bath scenario.

In the case of OBC and baths placed at the edges of the chain
[see Fig.~\ref{fig:sb}(a)], one has for the couplings
\begin{subequations}
	\begin{align}
		\gamma_{L,k} &= J(\omega_k) \frac{2 \sin^2(k)}{N+1},\\ 
		\gamma_{R,k} &= J(\omega_k) \frac{2 \sin^2(kN)}{N+1}.
	\end{align}
\end{subequations}
Making use of the quantization condition on $k$, one can show
with a straightforward calculation that
Eqs.~\eqref{eq:Ckq-1}, \eqref{eq:theta-is-2}, and~\eqref{eq:theta-is-3}
continue to remain valid.


\section{Scaling of entropy in the Kitaev chain} \label{sec:kitaev-ent}

We now turn to the steady-state von Neumann entropy
in the Kitaev chain with PBC. The blocks of the Majorana correlation matrix
$\Gamma$ are reported in Eq.~\eqref{eq:gamma-is},
where we recognize a Toeplitz structure.
Given a subsystem $A$, we are interested in the matrix
$\lambda\mathds{1}-i\Gamma_A$ [cf.~Eq.~\eqref{eq:D-def}],
which is then also of the Toeplitz type. Let us define 
its symbol $g_\lambda(k)$ [cf.~\eqref{eq:gamma-is}] as 
\begin{equation} \label{eq:glambda}
	g_\lambda(k) =
	\begin{bmatrix}
		\lambda & -i \theta_{kk} e^{-i\xi(k)} \\
		i\theta_{kk}e^{i\xi(k)} & \lambda
	\end{bmatrix}.
\end{equation}
In contrast with the tight-binding chain, for which it was a scalar, 
now the symbol is a two-by-two matrix.  
At zero temperature, the asymptotic behavior in the large-$\ell$
limit of the determinant of the Toeplitz matrix~\eqref{eq:gamma-is}
has been obtained in Ref.~\cite{its2005}.
Here we are only interested in the logarithmic correction to the volume-law 
scaling of the von Neumann entropy. Thus, we can use the techniques of 
Refs.~\cite{ares2018entanglement,ares2019sublogarithmic}.
The idea is that since the logarithmic 
correction depends only on the singularities of the symbol, we are allowed to 
modify the latter, provided that we do not change its singularity structure. 
This eventually allows one to work with a scalar symbol.
The computation is rather technical and we leave the details to
App.~\ref{app:kitaev}. The result is analogous to the tight-binding case:
\begin{equation} \label{eq:kitaev-deco}
	S_A = \alpha \ell + \frac{c'(\Theta)}{3} \ln(\ell) + \mathcal{O}(1),
\end{equation}
where $\alpha$ is the same constant reported in~\eqref{eq:lin-coeff} and
$c'(\Theta)$ is half of the effective central charge of the tight-binding model,
provided $|h| = 2$:
\begin{equation}
	c'(\Theta) = \frac{c(\Theta)}{2}.
\end{equation}
This time we have
\begin{equation} \label{eq:table}
	\Theta = \begin{cases}
		\theta_{kk}\eval_{k=\pi} & h = 2, \\
		\theta_{kk}\eval_{k=0} & h = -2,
	\end{cases}
\end{equation}
in accordance with the discussion at the end of Sec.~\ref{sec:kitaev}.
Clearly, in the zero-temperature limit $\Theta \to 1$ we recover the well-known
central charge $c = 1/2$ of the critical Kitaev chain.

This result is valid for a general value of $\theta_{kk}$, and we can easily
specialize the expression of $\Theta$ to the steady state~\eqref{eq:statio} of
our master equation. Since with PBC the matrix $\phi_{nk}$ is equal to the
corresponding matrix for the tight-binding chain [compare Eq.~\eqref{eq:phi-xx}
with Eq.~\eqref{eq:phi-is}], one finds for $K_{kq}$, $\theta_{kk}$, and $\Theta$
the same quantities reported in Eqs.~\eqref{eq:prev}-\eqref{eq:theta-single}
and Eqs.~\eqref{eq:Ckq-1}-\eqref{eq:theta-is-3} for the single-bath and
two-bath geometries, respectively.
In the case $\mu_L = \mu_R = 0$ and $T_L = T_R = T$ one recovers the correlator
for the finite-temperature Kitaev (and Ising) chain~\cite{barouch1970statistical,
barouch1971statistical,barouch1971statisticalmechanics} at temperature $T$.
For $T \to 0$ one finds $\theta_{kk} \to 1$, thus recovering the
zero-temperature correlator of the Kitaev chain. Importantly, the presence of
$\mu_L$ and $\mu_R$ in~\eqref{eq:theta-is-2} implies that the statistical
ensemble describing the steady state is not the usual finite-temperature
ensemble of the Kitaev chain.


\section{Numerical results} \label{sec:numerics}

We now provide numerical benchmarks for the results of Sec.~\ref{sec:tb-ent}
and Sec.~\ref{sec:kitaev-ent}.
In particular, we numerically diagonalize the Majorana correlation
matrix~\eqref{eq:gamma-block-struct} for our models and then we use
its eigenvalues to directly evaluate entropic quantities using
Eq.~\eqref{eq:S-nu}. The results are then compared to what we derived
in an analytical way.
In Sec.~\ref{sec:vn} we overview the behavior of the subsystem
von Neumann entropy. Our main results are contained in Sec.~\ref{sec:mi},
where we discuss the scaling behavior of the steady-state mutual information
both for the tight-binding chain (Sec.~\ref{sec:tb-numerics})
and for the Kitaev chain (Sec.~\ref{sec:ki-numerics}). 
Our numerical results confirm a logarithmic scaling for the mutual information,
being in perfect agreement with the predictions of the previous sections.
Finally, in Sec.~\ref{sec:negativity} we briefly argue that logarithmic scaling
also occurs for the fermionic logarithmic negativity, thus suggesting that
the growth of the mutual information reflects a logarithmic entanglement growth.


\subsection{Volume-law scaling of von Neumann entropy} \label{sec:vn}

In the presence of the external baths, the steady-state 
von Neumann entropy exhibits a volume-law scaling as $\alpha\ell$, 
with $\ell$ being the size of the subsystem
and $\alpha$ being the constant reported in Eq.~\eqref{eq:lin-coeff}, which
equals the density of the von Neumann entropy of the full system.
In the absence of baths, the full system is in a pure state,
and its von Neumann entropy is zero ($\alpha=0$).
The volume-law scaling in the
open setting is due to the fact that the steady state is described by a 
finite-temperature-like statistical ensemble
\cite{Karevski2009,Guarnieri2019}.

Here we focus on the tight-binding chain with OBC and one thermal bath
[see Fig.~\ref{fig:sb}(a)]. Results for different boundary 
conditions and for the Kitaev chain are qualitatively similar and will not be
discussed.

\begin{figure}
	\centering
	\includegraphics[scale=0.7]{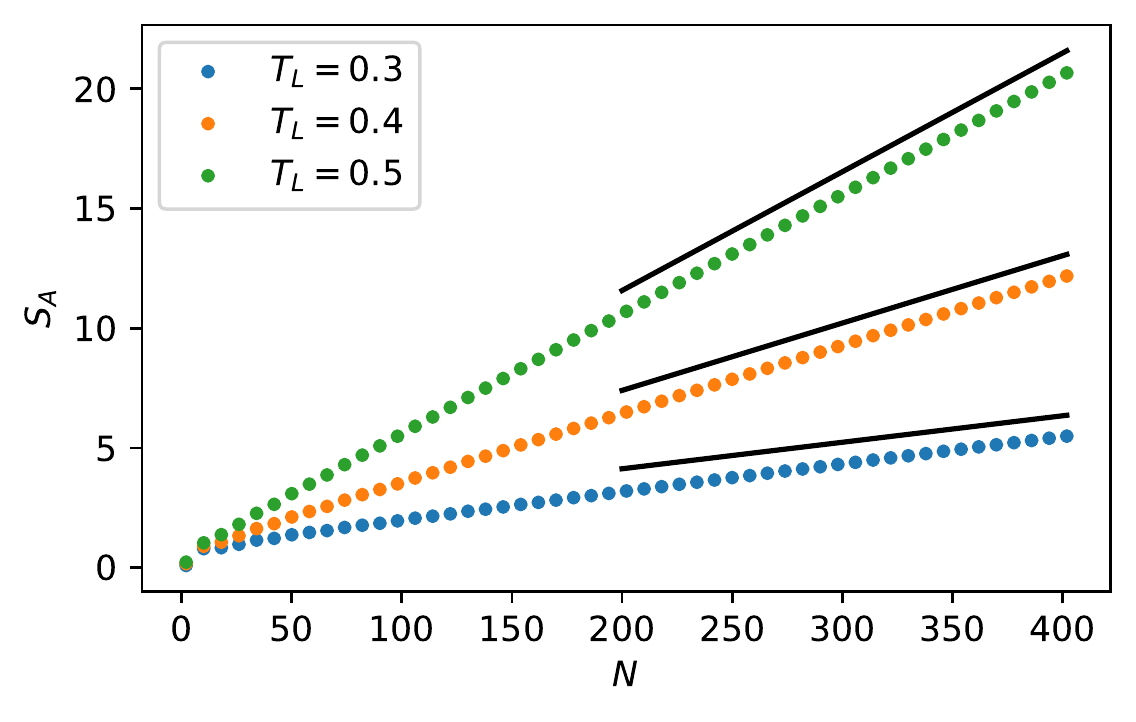}
	\caption{
		Volume-law scaling of the von Neumann entropy in the open
		tight-binding chain with a single bath on the left edge of the system
		(see Fig.~\ref{fig:sb}). Subsystem $A$ is the half chain ($\ell=N/2$). 
		Here we choose $h=1$, $\mu_L=-1$, and ${T_L=0.3,0.4,0.5}$. 
		Straight lines denote the analytic predictions for the volume-law 
		scaling, in the limit $\ell\to\infty$ [cf.~Eq.~\eqref{eq:ent-expansion}]. 
		The logarithmic correction is not clearly visible at this scale.
	}
	\label{fig:tb-volume}
\end{figure}

Figure~\ref{fig:tb-volume} reports the von Neumann entropy 
$S_A$ as a function of the chain length $N$, where the subsystem $A$
is the half-chain with $\ell = N/2$ [cf.~Fig.~\ref{fig:sb}(a)].
Only the left edge is in contact with a thermal bath at chemical potential
$\mu_L=-1$ and temperature $T_L$.
Data in the figure correspond to different temperatures $T_L$.
A robust growth is visible at all temperatures, with a slope that decreases
as the temperature is lowered. This is expected, since at $T=0$ the scaling 
of the von Neumann entropy is logarithmic with the interval size.
Continuous lines show the expected linear behavior as 
$\alpha\ell$ in the limit $\ell\to\infty$, with a prefactor 
$\alpha$ given by Eq.~\eqref{eq:lin-coeff}.
We observe a qualitative agreement with the theoretical predictions, at least
to the leading order in $\ell$. However, as also expected
from Eq.~\eqref{eq:ent-expansion}, subleading logarithmic corrections
are present. To reveal them it is convenient to use the mutual information.


\subsection{Logarithmic scaling of mutual information} \label{sec:mi}

We now discuss the scaling of the steady-state mutual information in the
presence of external baths. The logarithmic prefactor is determined by the
singular structure of the single-particle energy dispersion,
as discussed in Sec.~\ref{sec:tb-ent} and Sec.~\ref{sec:kitaev-ent}.


\subsubsection{Tight-binding chain} \label{sec:tb-numerics}

\begin{figure}
	\centering
	\includegraphics[scale=0.7]{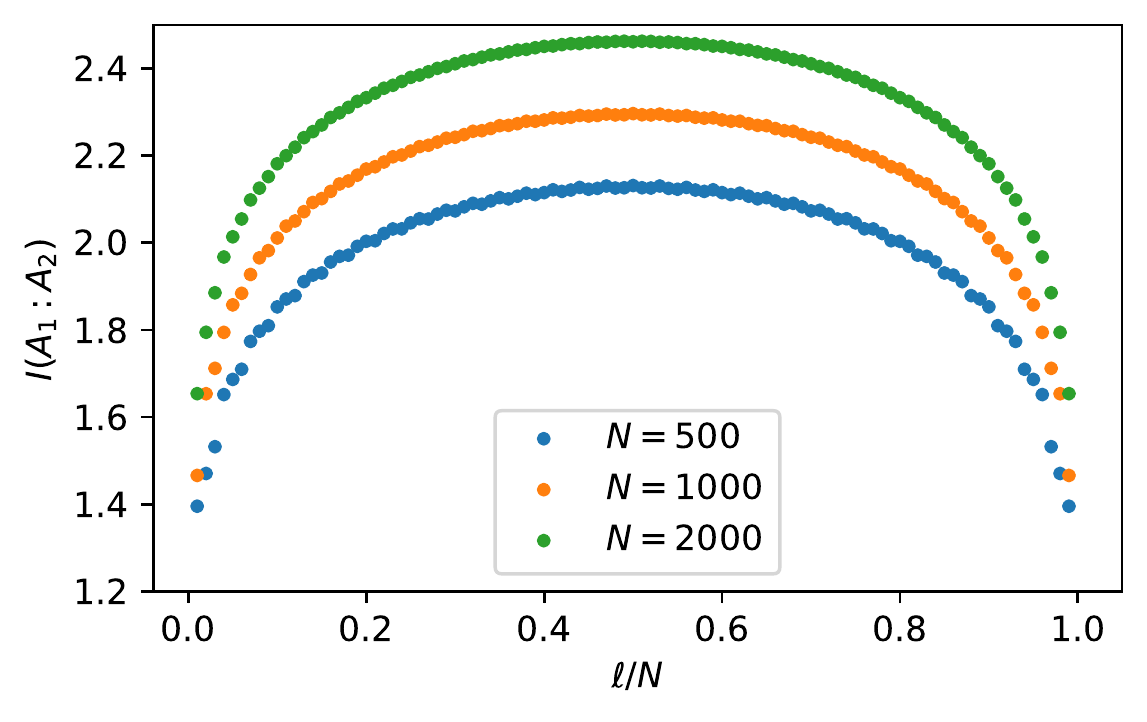}
	\caption{
		Mutual information $I(A_1:A_2)$ between two intervals 
		in the open tight-binding chain with a thermal bath on the left edge. 
		Here we choose $\mu_L=-1$, $T_L=0.3$, and $h=1$.
		Different colors correspond to different chain length $N$.
		The data are plotted versus $\ell/N$, 
		with $\ell$ being the size of $A_1$ (see Fig.~\ref{fig:sb}(a)). Notice 
		the symmetry under exchange of the two subsystems
		$\ell \leftrightarrow N-\ell$.
	}
	\label{fig:tb-single-no-scaling}
\end{figure}

For the tight-binding model of Eq.~\eqref{eq:ham-xx}, we consider the same setup
as in Sec.~\ref{sec:xx-single}, i.e, the open chain with a thermal bath
on the left edge. We fix $h=1$, $\mu_L=-1$, and $T_L=0.3$.
Our numerical data for the mutual information $I(A_1:A_2)$
between two complementary intervals 
$A_1$ and $A_2=\overline{A}_1$ are plotted in Fig.~\ref{fig:tb-single-no-scaling}
versus $\ell/N$, with $\ell$ being the size of $A_1$. The three different 
data sets correspond to different values of $N$. At each fixed $N$, $I(A_1:A_2)$ 
increases upon increasing $\ell$ up to $\ell\sim N/2$, after which it starts 
decreasing. The behavior at intermediate $1\ll\ell\ll N$ is consistent with 
a logarithmic increase, as predicted in Eq.~\eqref{eq:ent-expansion}, 
which should hold in the limit $N\to\infty$ and then
$\ell\to\infty$ (with this order of limits).

Looking now at the definition of the mutual information
[cf.~Eq.~\eqref{eq:mi-intro}], it is clear that, when constructing $I(A_1:A_2)$,
the volume-law terms in the entropies cancel out. 
To derive the prefactor of the logarithmic scaling, we can use
Eq.~\eqref{eq:ent-expansion} for each term appearing in~\eqref{eq:mi-intro}. 
Notice that no logarithmic contribution is expected from $S_{A_1\cup A_2}$,
since the entropy of the full system for large $N$ is exactly $\alpha N$,
with $\alpha$ given by~\eqref{eq:lin-coeff}. 
Let us also stress that, in principle, we are not allowed to use
Eq.~\eqref{eq:ent-expansion} for $S_{A_2}$ because the size $N-\ell$ of $A_2$ is 
comparable with $N$.
To proceed, we should then conjecture a generalization for an interval $A$ of
generic size, embedded in a finite-size chain. Following the standard strategy
for critical systems described by CFTs, we write~\cite{calabrese2009entanglement}
\begin{equation} \label{eq:A1-1}
	S_{A}=\alpha\ell+\frac{c(\Theta)}{6}\ln\left[\frac{N}{\pi}
	\sin\left(\frac{\pi\ell}{N}\right)\right]+{\mathcal O}(1).
\end{equation}
Notice that the prefactor of the volume-law term is the same as before,
while in the logarithmic term of Eq.~\eqref{eq:ent-expansion} we replaced
\begin{equation} \label{eq:chord}
	\ell \to X_\ell,\quad\mathrm{with}\,\,X_\ell \coloneqq
	\frac{N}{\pi}\sin\left(\frac{\pi\ell}{N}\right), 
\end{equation}
where $X_\ell$ is the so-called chord length.
Eq.~\eqref{eq:A1-1} holds in the thermodynamic limit $\ell,N\to\infty$.
For systems with boundaries, as is the case here, the actual chord length
differs from~\eqref{eq:chord} by an overall factor $2$, which only affects the
${\mathcal O}(1)$ term, and can therefore be neglected.
We conclude that
\begin{equation} \label{eq:theory-xx}
	I(A_1:A_2)= \frac{c(\Theta)}{3}\ln\left[\frac{N}{\pi}
	\sin\left(\frac{\pi\ell}{N}\right)\right]+{\mathcal O}(1). 
\end{equation}
The factor $1/3$ rather than $1/6$ is due to the fact that both subsystems $A_1$
and $A_2$ contribute with a logarithmic term. 
Importantly, Eq.~\eqref{eq:theory-xx} implies that for large $\ell,N$ the 
data for the mutual information should collapse on the same curve, when plotted 
as a function of $X_\ell$.

\begin{figure}
	\centering
	\includegraphics[scale=0.7]{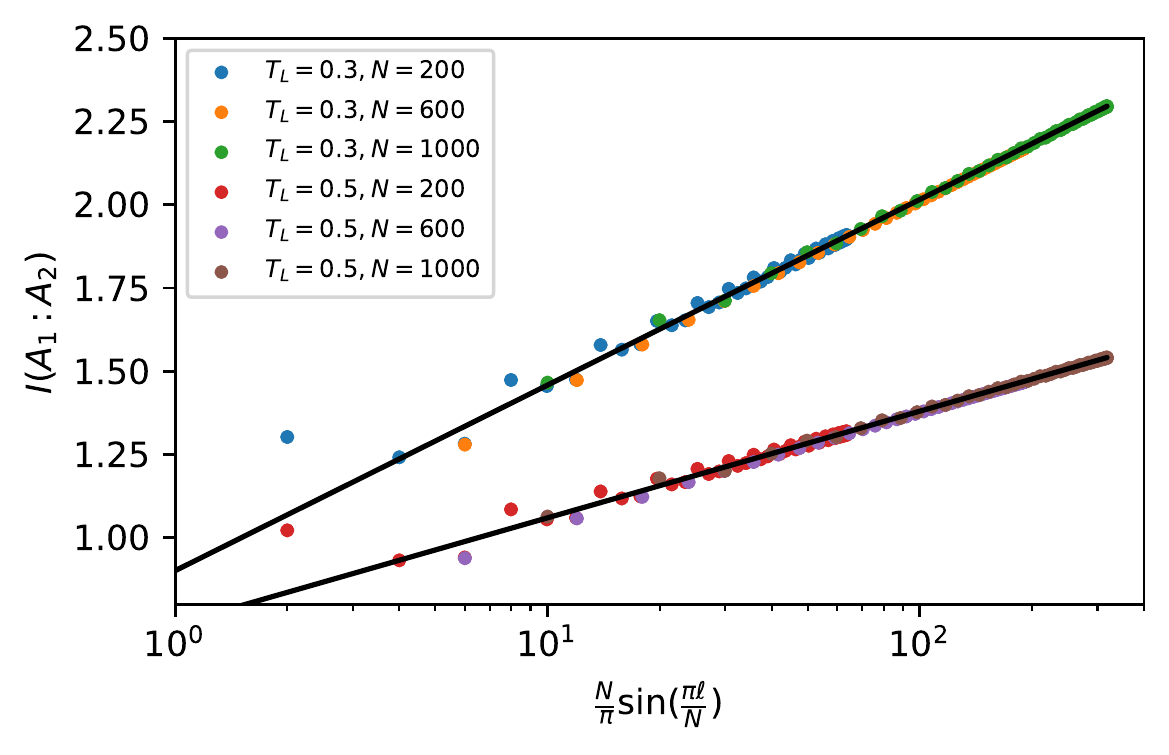}
	\caption{
		Mutual information $I(A_1:A_2)$ between two complementary 
		intervals [see Fig.~\ref{fig:sb}(a)] in the open tight-binding chain 
		with a thermal bath on the left edge. On the $x$-axis 
		$N\sin(\pi\ell/N)/\pi \equiv X_\ell$ is the chord length.
		Data are for fixed $\mu_L = -1$, $h=1$, and for $T_L=0.3,0.5$,
		while the various colors denote different $N$.
		The lines are fits to $c(\Theta)\ln(X_\ell)/3 + b$, 
		$c(\Theta)$ being the effective central charge
		[cf.~Eq.~\eqref{eq:ctheta-int}] and $b$ a fitting constant parameter.
	}
	\label{fig:tb-single-scaling}
\end{figure}

The validity of Eq.~\eqref{eq:theory-xx} is investigated in
Fig.~\ref{fig:tb-single-scaling} for the tight-binding chain with one thermal
bath on the left edge. We consider two different temperatures ${T_L=0.3,0.5}$
at fixed $\mu_L=-1$. The mutual information $I(A_1:A_2)$ is 
plotted versus $X_\ell$ (notice the logarithmic scale on the $x$-axis) 
for several values of ${N=200,600,1000}$. For both temperatures, the data 
exhibit collapse. The quality of the collapse improves upon increasing $N$,
as expected. Continuous lines are fits to Eq.~\eqref{eq:theory-xx},
where $c(\Theta)$ is kept fixed and given by Eq.~\eqref{eq:ctheta-int},
while the additive ${\mathcal O}(1)$ term being the only fitting 
parameter. For both temperatures, the agreement between the data and the fits
is very satisfactory.

\begin{figure}
	\centering
	\includegraphics[scale=0.7]{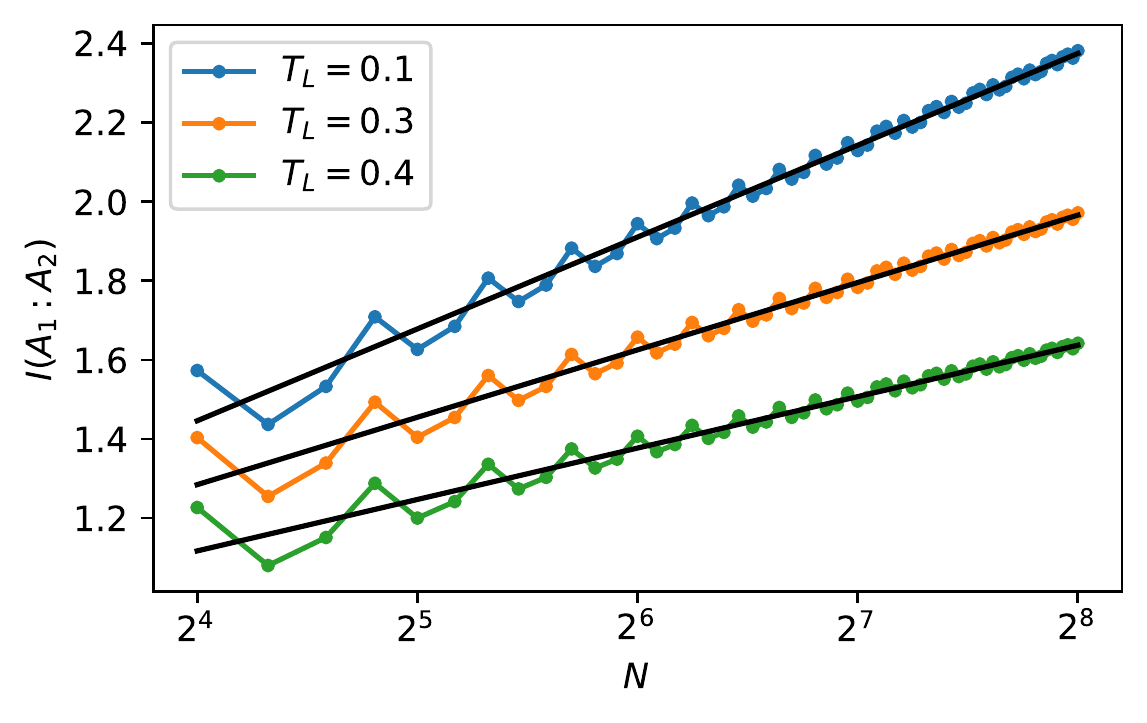}
	\caption{
		Scaling of the half-chain mutual information in 
		the open tight-binding chain with a thermal bath of the left edge 
		of the chain. Here we fix $\mu_L=-1$, $h=1$. Different 
		colors are for different values of $T_L$. Continuous lines 
		are fits to $a \ln(N)+b$, with $a,b$ fitting parameters.
	}
	\label{fig:tb-fits}
\end{figure}

\begin{figure}
	\centering
    \includegraphics[scale=0.7]{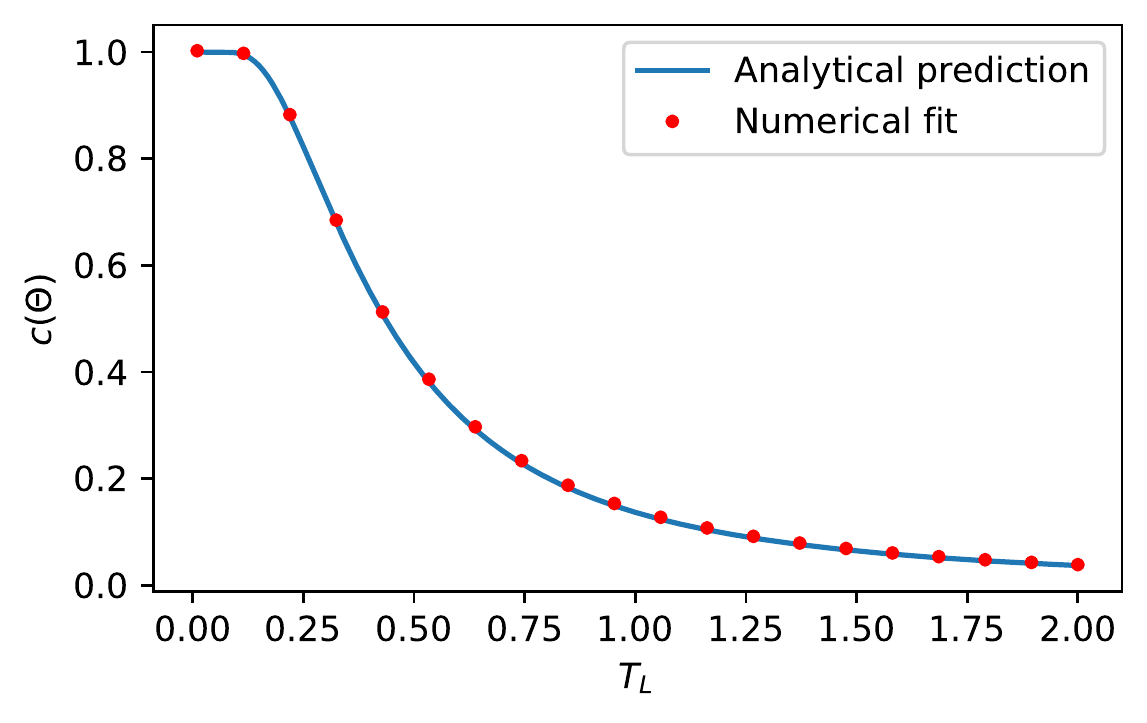}
    \caption{
		The effective central charge $c(\Theta)$ versus the 
		temperature $T_L$, as obtained from fits of numerical data
		as those in Fig.~\ref{fig:tb-fits}.
		Parameters are the same as in Fig.~\ref{fig:tb-fits}.
	}
    \label{fig:tb-check-single}
\end{figure}

Eq.~\eqref{eq:theory-xx} also implies that, for large $N$,
the mutual information between the two halves of the chain scales
logarithmically as $c(\Theta)\ln(N)/3$. This is shown 
in Fig.~\ref{fig:tb-fits}, for the same setup as in
Fig.~\ref{fig:tb-single-scaling}. 
The various data sets correspond to different temperatures of the external bath. 
The logarithmic increase is clearly visible (notice the semilog scale),
although oscillating corrections are present. The continuous lines are fits to 
\begin{equation} \label{eq:mi-fit}
	I(A_1:A_2) = a\ln(N)+b, 
\end{equation}
with $a,b$ fitting parameters.
Further checks of our results are provided in Fig.~\ref{fig:tb-check-single}
where, for each temperature, we numerically extract $c(\Theta)$ by fitting
the mutual information to Eq.~\eqref{eq:mi-fit}.
Symbols are the results of the fits, which are obtained as in
Fig.~\ref{fig:tb-fits} fitting the data with $N > 2^6$.
At low temperature, one finds $c(\Theta)\to1$, 
whereas $c(\Theta)$ vanishes in the high-temperature limit.
The continuous line is the analytic prediction in the limit $N\to\infty$,
given by Eq.~\eqref{eq:ctheta-int}. The agreement with the numerics is excellent.

\begin{figure}
	\centering
	\includegraphics[width=\columnwidth]{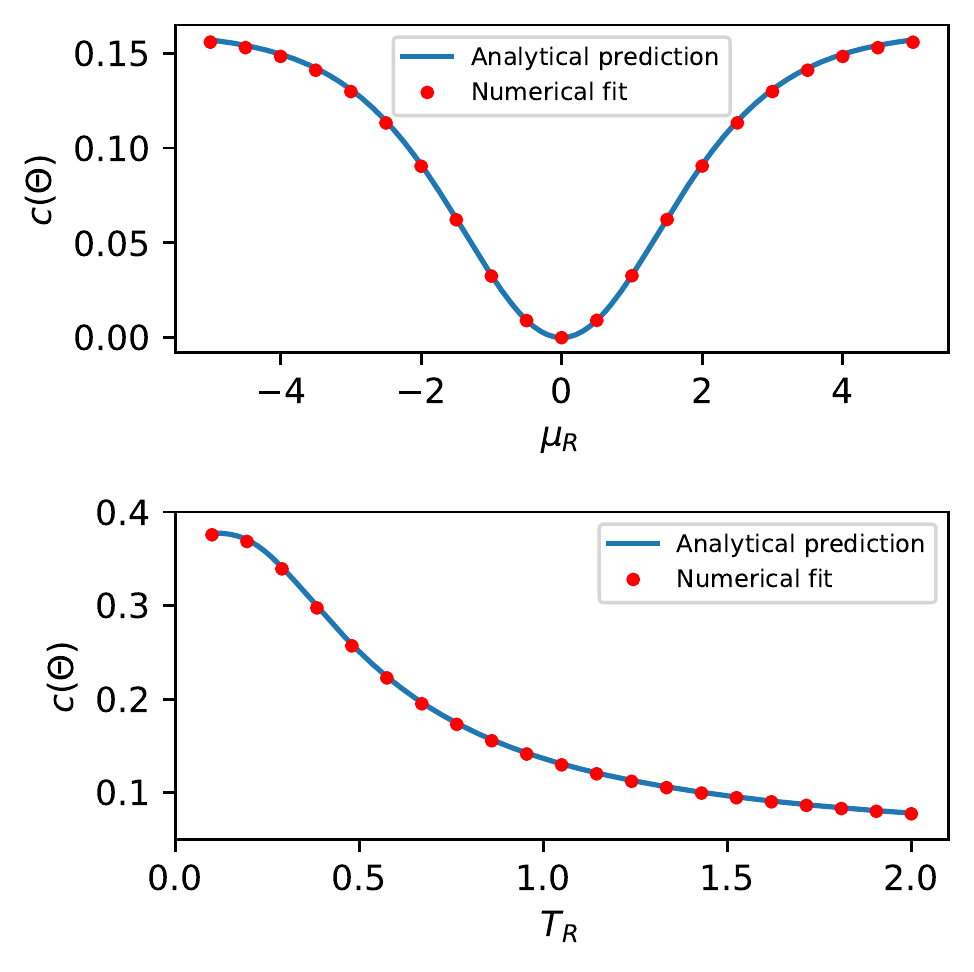}
	\caption{
		Logarithmic scaling of the mutual information $I(A_1:A_2)$ in the 
		tight-binding chain with OBC coupled to two 
		thermal baths at the edges. 
		In the upper panel, we plot $c(\Theta)$ versus $\mu_R$, for fixed
		$T_L = T_R = 1$ and $\mu_L = 0$. In the lower panel, we plot
		$c(\Theta)$ versus $T_R$, for fixed ${\mu_L = \mu_R = -1}$ and
		$T_L = 1$. In both cases, $h=1$.
		Numerical results for $c(\Theta)$ are obtained by performing
		a finite-size scaling analysis for the half-chain mutual information.
	}
	\label{fig:mi-tb-check-double}
\end{figure}

Finally we discuss a two-bath geometry, where the edges of the chain are
connected to two different thermal baths.
We fix $h=1$ and we consider two situations: constant temperature
$T_L = T_R = 1$ with $\mu_L = 0$ fixed, and constant chemical potential
$\mu_L = \mu_R = -1$ with $T_L = 1$ fixed. In Fig.~\ref{fig:mi-tb-check-double}
we plot the numerically extracted $c(\Theta)$ versus $\mu_R$ and $T_R$,
respectively, for the two scenarios.
As for Fig.~\ref{fig:tb-check-single},
the continuous line denotes the theoretical result in the limit $N \to \infty$,
which is in perfect agreement with the numerics.

\subsubsection{Kitaev chain} \label{sec:ki-numerics}

Let us now discuss the steady-state mutual information in the Kitaev chain. 
Here we consider a PBC geometry, as depicted in Fig.~\ref{fig:sb}(b).
Two sites at mutual distance $N/2$ are put in contact with two external 
baths at temperatures $T_{R/L}$ and with chemical potentials $\mu_{R/L}$. 
We choose $T_L=T_R=1$, $\mu_L=0$, $\mu_R=2,4$, and fix $h=2$.
We consider the mutual information $I(A_1:A_2)$ between two intervals of
size $\ell$ and $N/2-\ell$ placed between the baths (see Fig.~\ref{fig:sb}).
First, we should observe that 
in constructing the mutual information~\eqref{eq:mi-intro} all the entropies
(i.e., $S_{A_1}$, $S_{A_2}$, and $S_{A_1\cup A_2}$) contain a subleading
logarithmic term. This happens because $S_{A_1\cup A_2}$ is not the full system.
As for the tight-binding chain the volume-law terms, instead, cancel out.
The final result is 
\begin{equation} \label{eq:theory-kit}
	I(A_1:A_2) = \frac{c'(\Theta)}{3}\ln(X_{2\ell}) + {\mathcal O}(1),
\end{equation}
where $X_{2\ell}$ is the chord length in Eq.~\eqref{eq:chord}
(notice the factor $2$), and $c'(\Theta) = c(\Theta)/2$ is the effective
central charge calculated for the Kitaev chain [cf.~Eq.~\eqref{eq:kitaev-deco}].
To derive Eq.~\eqref{eq:theory-kit} we used the fact that,
for all the intervals $A_1$, $A_2$, and $A_1\cup A_2$,
\begin{equation} \label{eq:Sw}
	S_{W}\xrightarrow[]{\ell,N\to\infty}\frac{c'(\Theta)}{3}
	\ln(X_\ell), \quad W=A_{1(2)},A_1\cup A_2, 
\end{equation}
After substituting~\eqref{eq:Sw} in the definition of the 
mutual information~\eqref{eq:mi-intro}, we obtain~\eqref{eq:theory-kit}. 
We point out that Eq.~\eqref{eq:theory-kit} holds only for the geometry 
in Fig.~\ref{fig:sb}(b), although it could be easily generalized to
more general settings.

\begin{figure}
	\centering
    \includegraphics[scale=0.7]{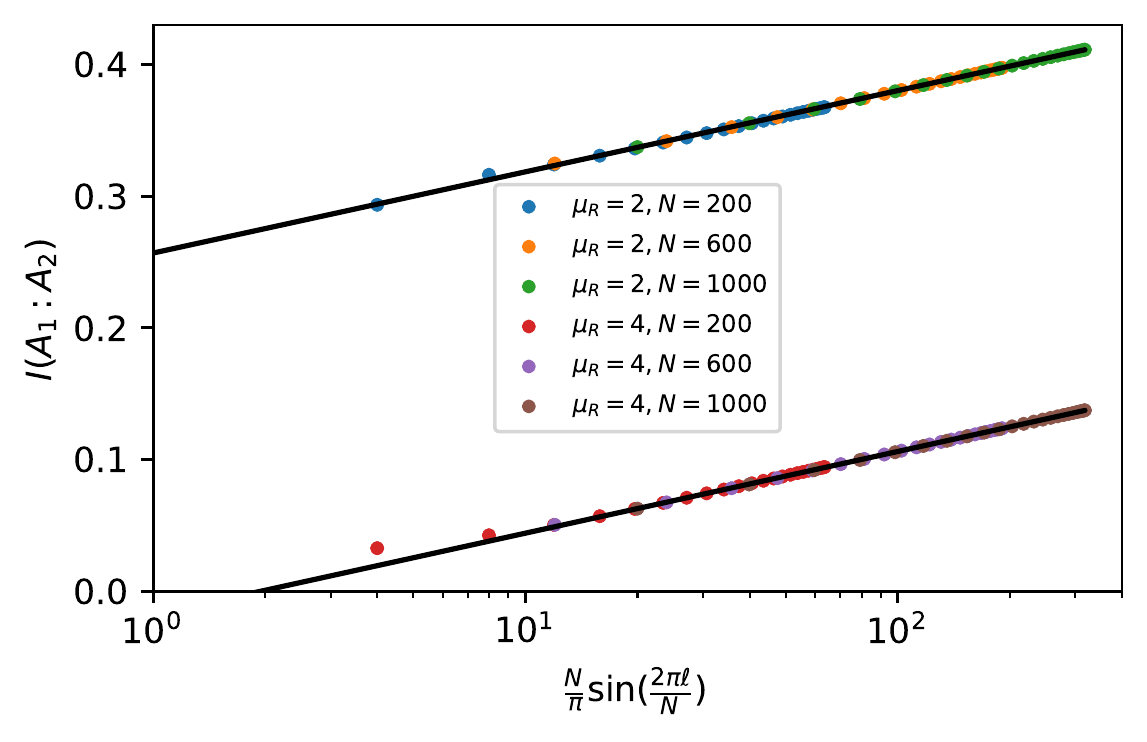}
    \caption{
		Mutual information $I(A_1:A_2)$ between two intervals in the Kitaev
		chain with PBC and two external baths [see Fig.~\ref{fig:sb}(b)] versus
		the chord length $X_{2\ell}=N\sin(2\pi\ell/N)/\pi$.
		Here we choose $T_L=T_R=1$, $\mu_L=0$ and $\mu_R=2,4$.
		The various symbols correspond to different chain sizes $N$.
		Continuous lines are fits to  
		$I(A_1:A_2)=c(\Theta)\ln(X_{2\ell})/6 + b$, with 
		$c(\Theta)$ the effective central charge, and $b$ a fitting constant.
	}
    \label{fig:kitaev}
\end{figure}

The validity of Eq.~\eqref{eq:theory-kit} is numerically verified in
Fig.~\ref{fig:kitaev}, where we plot $I(A_1:A_2)$ versus $X_{2\ell}$.
For both values of $\mu_R$, the data exhibit collapse at large $\ell, N$.
Continuous lines are fits to Eq.~\eqref{eq:theory-kit}, 
the only fitting parameter being the ${\mathcal O}(1)$ constant.
The agreement between the analytic prediction in the scaling limit 
$\ell,N\to\infty$ and the numerics is nearly perfect already for
relatively small chains with $X_{2\ell}\sim 10$.

\subsection{Fermionic logarithmic negativity} \label{sec:negativity}

In general, the mutual information between two subsystems does not provide
a measure of quantum entanglement between them, since it contains information
also about classical correlations. Given $A = A_1 \cup A_2$ in a mixed state
$\rho_A$, a proper measure of entanglement between $A_1$ and $A_2$ is given by
the fermionic logarithmic negativity~\cite{shapourian2017,shapourian2019},
defined as
\begin{equation} \label{eq:neg}
	\mathcal{E}(A_1 : A_2) = \ln\norm{\rho_A^{R_1}},
\end{equation}
where $\norm{X} = \Tr\sqrt{X X^\dagger}$ is the trace norm and $\rho_A^{R_1}$
stands for the operator obtained after performing a partial time-reversal
transformation on $\rho_A$ with respect to $A_1$ (notice that this is different
from the standard logarithmic negativity defined for bosonic
systems~\cite{Vidal_2002}). The fermionic logarithmic negativity can be
efficiently calculated numerically for fermionic Gaussian
states~\cite{shapourian2017}, and this applies in particular to our steady state.

\begin{figure}
	\centering
    \includegraphics[scale=0.7]{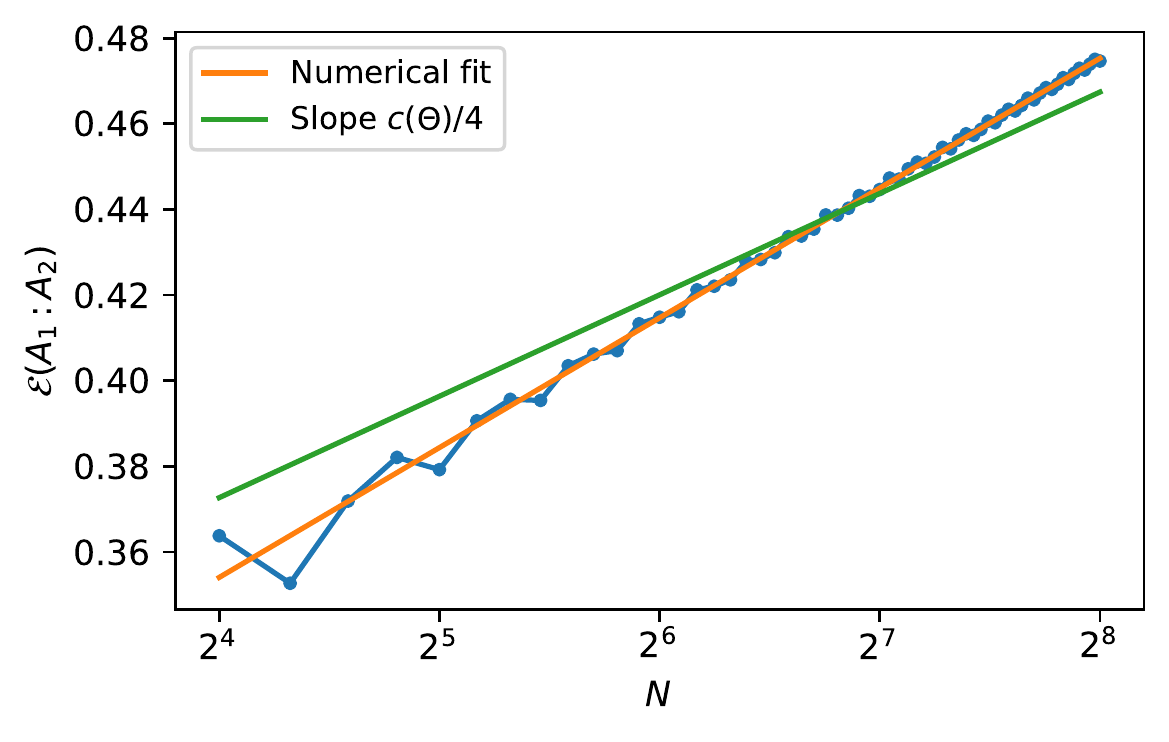}
    \caption{
		Scaling of the half-chain fermionic logarithmic
		negativity~\eqref{eq:neg} in the tight-binding chain with OBC and two
		thermal baths at the edges. Here the parameters are
		${h=1}$, ${\mu_L = -1}$, ${\mu_R = -1.5}$, ${T_L = 1}$, ${T_R = 1.5}$.
		The orange line is a linear fit in logarithmic scale, while the green
		line is $(c(\Theta)/4)\ln N + b$ with $b$ a fitting constant.
	}
    \label{fig:neg}
\end{figure}

Let us consider, as an example, the fermionic logarithmic negativity between
two halves of a tight-binding chain with OBC. In the ground state of the
isolated chain it is known that the fermionic logarithmic negativity exhibits
logarithmic scaling with a prefactor of $c/4$~\cite{shapourian2017}.
In Fig.~\ref{fig:neg} we report
an example of calculation with our steady state in the two-bath geometry.
We clearly observe logarithmic scaling, suggesting that such a feature
is of quantum nature. However, the prefactor is not consistent with a
straightforward generalization $c/4 \to c(\Theta)/4$, hence further studies
are necessary in order to establish its exact value in the
nonequilibrium scenario. Similar conclusions apply for the Kitaev chain.

\section{Conclusions} \label{sec:conclusions}

We investigated the quantum-information spreading in the tight-binding chain
and the Kitaev chain in the presence of external thermal baths coupled to
individual sites of the chains. To this purpose, we employed a self-consistent
\emph{nonlocal} Lindblad master equation approach, where the Lindblad operators 
modeling the baths are written in terms of the Bogoliubov modes that diagonalize
the isolated system, implying that they are, in principle, nonlocal
in real space~\cite{dabbruzzo2021self}.
The statistical ensemble describing the steady state is written in terms of a
convex combination of the Fermi-Dirac distributions of the baths. 
We showed that the steady-state von Neumann entropy of a subsystem exhibits a 
volume-law scaling with the subsystem size, reflecting that the system is not
in a pure state. The mutual information exhibits an area-law scaling for generic 
values of the system parameters. Interestingly, we observe logarithmic
violations of the area law in the presence of ground-state criticality. 
This behavior reflects the singularity of the single-particle 
energy dispersion of the models, which is present at all energies. 
We analytically derived the prefactor of the logarithmic growth of the mutual 
information, which depends on the system and bath parameters,
such as the temperature and the chemical potential. 

Let us now mention some promising directions for future work.
First of all, here we only analyzed the steady-state value of the
mutual information: it would be tempting to study the full-time dynamics,
in order to establish how the logarithmic scaling builds up during the evolution
of the system. A natural conjecture is that the same effective central charge
governs a logarithmic increase in time, as in Ref.~\cite{Kormos2017}. 
Our analysis may be also extended to genuine quantum entanglement measures
for mixed states, such as the fermionic logarithmic negativity.
Finally, it would be important to check the validity of our results by
comparing them with \emph{ab initio} numerical simulations, or with results
obtained using different master equations. A crucial question to address
is whether the logarithmic scaling of the mutual information would survive
in interacting integrable systems (or even in nonintegrable ones),
or in the presence of non-Markovian interactions with the environment.


\appendix

\section{Tight-binding model in terms of Fourier modes}
\label{app:fourier}

In this appendix we argue that the presence of the discontinuity in the
symbol~\eqref{eq:gnm} is not artificially introduced by our choice of the basis
with which we diagonalized the tight-binding Hamiltonian~\eqref{eq:ham-xx},
that is performed with discontinuous coefficients~\eqref{eq:phipsi-xx}.
Specifically, let us consider a tight-binding chain with PBC and define Fourier
modes $c_k$ through
\begin{equation}
	a_n = \frac{1}{\sqrt{N}} \sum_k e^{-ikn} c_k.
\end{equation}
In the basis $\{c_k\}$ the Hamiltonian is diagonalized, as in~\eqref{eq:ek},
but with single-particle energies
\begin{equation}
	\omega_k = -h-2\cos(k).
\end{equation}
The master equation of Ref.~\cite{dabbruzzo2021self} can be derived in a
straightforward way to obtain Eq.~\eqref{eq:dissipator1}, but with the
substitution $b_k \to c_k$. However, Eq.~\eqref{eq:D-self} is no longer valid
because $\omega_k$ may be negative.
The calculation of the steady-state correlation functions can still be performed
starting directly from Eq.~\eqref{eq:dissipator1}. Using the relation
$\Gamma_\alpha(\omega) + \Gamma_\alpha(-\omega) = J_\alpha(|\omega|)$ we find
\begin{equation}
	\langle c_k^\dagger c_k \rangle = \frac{\sum_\alpha \Phi_{\alpha,k}
	\Gamma_\alpha(-\omega_k)}{\sum_\alpha \Phi_{\alpha,k}
	J_\alpha(|\omega_k|)} =
	\begin{cases}
		\langle b_k^\dagger b_k \rangle & \omega_k > 0, \\
		1 - \langle b_k^\dagger b_k \rangle & \omega_k < 0,
	\end{cases}
\end{equation}
where $\langle b_k^\dagger b_k \rangle$ is the standard Bogoliubov correlator 
reported in~\eqref{eq:statio} with $\omega_k \to |\omega_k|$. If the model is 
critical, then $\langle c_k^\dagger c_k \rangle$ is discontinuous
as a function of $k$.

If we now define the Majorana operators in terms of $c_k$, we obtain
[cf.~Eq.~\eqref{eq:gnm}]
\begin{equation}
	G_{nm} = \int_{-\pi}^\pi \frac{dk}{2\pi}
	(1-2\langle c_k^\dagger c_k \rangle) e^{ik(n-m)},
\end{equation}
which, in terms of $\langle b_k^\dagger b_k \rangle$, becomes
\begin{equation}
	G_{nm} = \int_{-\pi}^\pi \frac{dk}{2\pi}
	\mathrm{sgn}(\omega_k) (1-2\langle b_k^\dagger b_k \rangle) e^{ik(n-m)},
\end{equation}
which is identical to Eq.~\eqref{eq:gnm}. Therefore, the von Neumann entropy is 
unaltered by this change of basis, as expected.


\section{Calculation for the tight-binding chain}
\label{app:tight-binding}

In this appendix we show how to perform the calculation of the steady-state
von Neumann entropy for the tight-binding chain (cf.~Sec.~\ref{sec:tb-ent}).

Let us first consider the case of PBC, i.e., $\zeta=0$ in Eq.~\eqref{eq:gnm}.
From that equation we obtain  
\begin{equation} \label{eq:matg}
	\lambda\delta_{nm}-G_{nm}=
	\int_{-\pi}^\pi\frac{dk}{2\pi} e^{ik(n-m)} g_\lambda(k).
\end{equation}
This defines a Toeplitz matrix, hence we can apply the Fisher-Hartwig
theorem~\cite{fisher1969toeplitz,basor1991the,basor1994the} to evaluate its
determinant for large $\ell$. Such theorem has been already employed in the
literature to determine the scaling behavior of the von Neumann and R\'enyi
entropies in the ground state of critical fermionic
chains~\cite{jin2004quantum,calabrese2010universal,fagotti2011universal}.
Here we apply a specialized version in which the symbol $g_\lambda(k)$
is allowed to have only jump discontinuities at a finite number of points $k_r$.
In order to apply the theorem one has to rewrite $g_\lambda(k)$ in the form
\begin{equation} \label{eq:symbol}
	g_\lambda(k) = g_s(k) \prod_{r=1}^R e^{ib_r( k-k_r-\pi \,
	\mathrm{sgn}(k-k_r) )}.
\end{equation}
Here $g_s(k)$ is a smooth function of $k$, $R$ is the number of discontinuities
of the symbol, and $b_r,k_r$ are real constants. The Fisher-Hartwig theorem 
states that, in the limit $\ell \to \infty$, one has 
\begin{equation} \label{eq:D-asy}
	D_\ell[g_\lambda] \sim F[g_s]^\ell
	\Bigg(\prod_{j=1}^R\ell^{-b_j^2} \Bigg) E[g_\lambda],
\end{equation}
where we defined 
\begin{equation} \label{eq:F-capital}
	F[g_s] \coloneqq \exp\left(\int_{-\pi}^\pi\frac{dk}{2\pi}\ln g_s(k)\right). 
\end{equation}
From Eq.~\eqref{eq:ent-C} it is clear that the first factor in~\eqref{eq:D-asy}
gives a volume-law von Neumann entropy, and it is not sensitive to the 
singularities in the symbol $g_\lambda(k)$. The second factor is responsible for 
the logarithmic scaling of the von Neumann entropy and contains information
about the singularities of $g_\lambda(k)$. The constant $E$ is a known function
of $g_\lambda(k)$. In the following, we are not considering $E$ because
we are interested only in the linear growth of the von Neumann entropy and 
in the logarithmic correction.
 
It is straightforward to check that in our case the symbol $g_\lambda(k)$ 
in~\eqref{eq:tilde-simb} can be written in the form~\eqref{eq:symbol} 
with two discontinuities at $k_1=-k_F$ and $k_2=k_F$, i.e., with $R=2$, and 
\begin{equation} \label{eq:b}
	b_1 = -b_2 = \beta_\lambda + m,
\end{equation}
where $m$ is an integer and 
\begin{equation}\label{eq:beta}
	\beta_\lambda = \frac{1}{2\pi i}\ln\left(\frac{\lambda-
	\Theta}{\lambda+\Theta}\right), \quad 
	\Theta \coloneqq \theta_{kk}\eval_{k=k_F}. 
\end{equation}
The function $g_s(k)$ is given by
\begin{multline} \label{eq:f}
	g_s(k)=\left(\frac{\lambda+\Theta}{\lambda-\Theta}\right)^{\frac{k_F}{\pi}-1}
	(\lambda+\theta_{kk})\Theta_{\mathrm H}(k_F-|k|)\\
	+ \left(\frac{\lambda+\Theta}{\lambda-\Theta}\right)^{\frac{k_F}{\pi}}
	(\lambda-\theta_{kk})\Theta_{\mathrm H}(|k|-k_F). 
\end{multline}
Now, we have to substitute Eqs.~\eqref{eq:b}, \eqref{eq:beta}, and~\eqref{eq:f} 
in Eq.~\eqref{eq:D-asy}.
The first factor in Eq.~\eqref{eq:D-asy} determines the constant $\alpha$ in
Eq.~\eqref{eq:ent-expansion}. By using~\eqref{eq:new-contour} one obtains
\begin{equation} \label{eq:line-inte}
	\alpha \! = \! 
	\lim_{\delta,\epsilon \to 0^+} \frac{1}{4\pi^2 i}
	\oint_\gamma d\lambda\int_{-\pi}^\pi dk \frac{e(1+\epsilon,\lambda)}
	{\lambda+\mathrm{sgn}(k_F-|k|)\theta_{kk}}, 
\end{equation}
where $e(x,\nu)$ is defined in~\eqref{eq:e-def}, $\theta_{kk}$
in~\eqref{eq:theta-def}, and $\gamma$ denotes the contour shown in
Fig.~\ref{fig:contour}. This integral can be performed with the residue theorem,
leading to
\begin{equation}
	\alpha = \int_{-\pi}^\pi \frac{dk}{2\pi} e(1,\theta_{kk}),
\end{equation}
which is precisely the expression reported in Eq.~\eqref{eq:lin-coeff}.

The second factor in Eq.~\eqref{eq:D-asy} yields for the prefactor of the
logarithmic term (cf.~Eq.~\eqref{eq:ent-expansion} with $\nu = 1$)
\begin{equation} \label{eq:ctheta}
	\frac{c(\Theta)}{3} = \lim_{\delta,\epsilon \rightarrow 0^+} \frac{1}{\pi i}
	\oint_\gamma d\lambda \,
	e(1+\epsilon,\lambda) \frac{d(-\beta_\lambda^2)}{d\lambda},
\end{equation}
where $\beta_\lambda$ is defined in Eq.~\eqref{eq:beta} and,
again, $\gamma$ denotes the ``dogbone'' contour in Fig.~\ref{fig:contour}. 
We can perform an integration by parts to obtain 
\begin{equation} \label{eq:dogbone}
	\frac{c(\Theta)}{3}= 
    \lim_{\delta,\epsilon \rightarrow 0^+} \frac{1}{8\pi^3 i}
    \oint_\gamma d\lambda \, \ln^2 \left( \frac{\lambda+\Theta}
    {\lambda-\Theta} \right) \ln(\frac{1+\epsilon+\lambda}
    {1+\epsilon-\lambda}).
\end{equation}
The contribution of the circles around $\pm 1$ in $\gamma$
(see Fig.~\ref{fig:contour}) vanishes in the limit $\epsilon \to 0^+$.
The integration along the horizontal paths can be performed using the fact that, 
for $\delta \to 0^+$, one has  
\begin{equation} \label{eq:id-delta}
	\ln(\frac{x \pm i\delta + t}{x \pm i \delta - t})
	\to \ln\left|\frac{t + x}{t - x}\right| \mp i\pi\, 
	\text{sgn}(t)\Theta_{\mathrm H}(|t|-|x|).
\end{equation}
Inserting in~\eqref{eq:dogbone}, this gives
\begin{equation} \label{eq:ctheta-integral}
	\frac{c(\Theta)}{3} = \frac{1}{2\pi^2} \int_{-\Theta}^\Theta dx
	\ln(\frac{\Theta+x}{\Theta-x}) \ln(\frac{1+x}{1-x}).
\end{equation}
This integral can be expressed in terms of dilogarithm
functions~\eqref{eq:dilog}: the result is reported in Eq.~\eqref{eq:ctheta-int}.

Let us now discuss the case with OBC and consider a block 
of $\ell$ sites starting at one edge of the chain [see Fig.~\ref{fig:sb}(a)]. 
Now, one has $\zeta=1$ in the fermionic correlator~\eqref{eq:gnm}, which has the 
Toeplitz-plus-Hankel structure with the same symbol.
A version of the Fisher-Hartwig theorem for certain kinds of Toeplitz-plus-Hankel 
matrices exists~\cite{deifts2011asymptotics}, and the analysis can be carried out
in a similar way as before. In our particular scenario, we only have to change
$b_j^2 \to b_j^2/2$ in Eq.~\eqref{eq:D-asy}, which has the effect of halving
the coefficient of the logarithmic term~\cite{fagotti2011universal}.
This justifies the validity of Eq.~\eqref{eq:ent-expansion} with $\nu = 2$.


\section{Calculation for the Kitaev chain}
\label{app:kitaev}

In this appendix we show how to perform the calculation of the steady-state
von Neumann entropy for the Kitaev chain with PBC
(cf.~Sec.~\ref{sec:kitaev-ent}).

The starting point is the symbol reported in Eq.~\eqref{eq:glambda}.
In order to calculate the associated Toeplitz determinant, the idea is to
modify the symbol without altering its singularity structure, so that we can
reduce to a calculation with scalar
symbols~\cite{ares2018entanglement,ares2019sublogarithmic}
Let us define the modified symbol $\widetilde g_\lambda(k)$ as 
\begin{equation} \label{eq:gtilde}
	\widetilde{g}_\lambda(k) \coloneqq
	\begin{bmatrix}
		\lambda & -i\theta_{kk} e^{-i|\xi(k)|} \\
		i\theta_{kk} e^{i|\xi(k)|} & \lambda
	\end{bmatrix},
\end{equation}
which differs from $g_\lambda(k)$ because of the absolute value $|\xi(k)|$
in the phase factors. Its inverse $\widetilde g^{-1}_\lambda(k)$ is 
\begin{equation} \label{eq:gtilde-i}
	\widetilde{g}_\lambda^{-1}(k) =
	\frac{1}{\lambda^2 - \theta_{kk}^2}
	\begin{bmatrix}
		\lambda & i\theta_{kk} e^{-i|\xi(k)|} \\
		-i\theta_{kk} e^{i|\xi(k)|} & \lambda
	\end{bmatrix}.
\end{equation}
Crucially, both $\widetilde g_\lambda$ and its inverse are smooth functions 
of $k$. As a consequence, in the limit $\ell\to\infty$,
the corresponding Toeplitz determinants
$D_\ell[\widetilde g_\lambda]$ and $D_\ell[\widetilde g^{-1}_\lambda]$ 
are not expected to contain logarithmic terms. 
Their asymptotic behavior is in fact determined  
by the Szeg\"o-Widom theorem~\cite{widom1976asymptotic}. Given a generic 
smooth symbol $z(k)$ of a Toeplitz matrix, the Szeg\"o-Widom theorem gives 
\begin{equation} \label{eq:widom}
	\ln D_\ell[z] = \ell \int_{-\pi}^\pi \frac{dk}{2\pi} \ln\det(z(k))
	+ \mathcal{O}(1).
\end{equation}
In our case, $z(k)=\widetilde g_\lambda(k)$ or
$z(k) = \widetilde g^{-1}_\lambda(k)$.
Moreover, we can use the so-called Basor localization theorem~\cite{basor1979a},
which allows us to write, in the limit $\ell \to \infty$,
\begin{equation} \label{eq:Dlambda}
	\ln D_\ell[{g}_\lambda] = \ln D_\ell[{g}_\lambda 
	\widetilde{g}_\lambda^{-1}] -
	\ln D_\ell[\widetilde{g}_\lambda^{-1}] + \mathcal{O}(1). 
\end{equation}
Here the first contribution contains logarithmic terms, 
whereas the second one gives rise to volume-law terms as in~\eqref{eq:widom}. 
To proceed, let us now notice that
\begin{equation} \label{eq:ggtilde}
	g_\lambda \widetilde{g}_\lambda^{-1} \!= \!
	\frac{1}{\lambda^2 - \theta_{kk}^2}
	\begin{bmatrix}
		\lambda^2 - \theta_{kk}^2 e^{i(|\xi|-\xi)} &
		i\lambda\theta_{kk} (e^{-i|\xi|} \!-\! e^{-i\xi}) \\
		i\lambda\theta_{kk} (e^{i\xi} \!-\! e^{i|\xi|}) &
		\lambda^2 - \theta_{kk}^2 e^{i(\xi - |\xi|)}
	\end{bmatrix}\!.
\end{equation}
If $-\pi \leq k \leq 0$, then $\xi(k) \geq 0$ and
$g_\lambda\widetilde g_\lambda^{-1}$ is the identity matrix.
On the other hand, if $0 < k \leq \pi$, then $\xi(k) < 0$
and Eq.~\eqref{eq:ggtilde} becomes 
\begin{equation} \label{eq:ggtilde-1}
	g_\lambda \widetilde{g}_\lambda^{-1} =
	\frac{1}{\lambda^2 - \theta_{kk}^2}
	\begin{bmatrix}
		\lambda^2 - \theta_{kk}^2 e^{-2i\xi} &
		-2\lambda\theta_{kk} \sin(\xi) \\
		-2\lambda\theta_{kk} \sin(\xi) &
		\lambda^2 - \theta_{kk}^2 e^{2i\xi}
	\end{bmatrix}.
\end{equation}
Diagonalizing the matrix in~\eqref{eq:ggtilde-1}, one obtains that the
eigenvalues $b_\pm$ are
\begin{equation} \label{eq:is-eig}
	b_\pm = \left[ \frac{\sqrt{\lambda^2 - \theta_{kk}^2 \cos^2(\xi)} \pm
	|\theta_{kk} \sin(\xi)|}
	{\sqrt{\lambda^2 - \theta_{kk}^2}} \right]^2 .
\end{equation}
One can easily verify that the corresponding eigenvectors are smooth 
functions of $k$. Hence, a further application of Basor localization
theorem yields 
\begin{equation}
	\label{eq:semi-final}
  \ln D_\ell[g_\lambda \widetilde{g}_\lambda^{-1}] =
  \ln D_\ell[b_-] + \ln D_\ell[b_+] + \mathcal{O}(1). 
\end{equation}
Here we also used that, according to the Szeg\"o-Widom theorem~\eqref{eq:widom},
the contribution in Eq.~\eqref{eq:semi-final} of the matrices that 
diagonalize~\eqref{eq:ggtilde-1} would be a constant that we can neglect
in the limit $\ell\to\infty$. Now in~\eqref{eq:semi-final}, 
$D_\ell[b_{\pm}]$ are determinants of Toeplitz matrices with scalar 
symbols. Their asymptotic behavior for large $\ell$ 
can be determined by using the standard Fisher-Hartwig theorem
(as in App.~\ref{app:tight-binding}). We obtain
\begin{multline} \label{eq:bpm-asy}
	\ln D_\ell[b_\pm] = \ell \int_{-\pi}^\pi \frac{dk}{2\pi} \, \ln b_\pm(k) \\
	+ \ln^2 \left[ \frac{\sqrt{\lambda^2}
	\pm |\Theta|}{\sqrt{\lambda^2 - \Theta^2}} \right]
	\frac{\ln(\ell)}{\pi^2} + \mathcal{O}(1),
\end{multline}
where $\Theta$ is reported in Eq.~\eqref{eq:table}.
Noticing that $b_- b_+ = 1$, and using Eqs.~\eqref{eq:widom},
\eqref{eq:Dlambda}, and~\eqref{eq:bpm-asy}, we get
\begin{multline} \label{eq:kitaev-fin}
	\ln D_\ell[g_\lambda] = 
	\ell \int_{-\pi}^\pi \frac{dk}{2\pi} \ln(\lambda^2 - \theta_{kk}^2) \\
	+ \ln^2\left[ \frac{\sqrt{\lambda^2} + |\Theta|}{\sqrt{\lambda^2 -
	\Theta^2}} \right] \frac{2\ln(\ell)}{\pi^2} + \mathcal{O}(1).
\end{multline}

Now we can determine the scaling of the steady-state von Neumann entropy
using Eq.~\eqref{eq:ent-C}. The first term in~\eqref{eq:kitaev-fin} leads to
the coefficient of the volume-law term $\alpha$, which turns out to be the same
as the tight-binding one~\eqref{eq:lin-coeff}. The second term
in~\eqref{eq:kitaev-fin} leads instead to [cf.~Eq.~\eqref{eq:kitaev-deco}]
\begin{equation}
	\frac{c'(\Theta)}{3} = \!\lim_{\delta,\epsilon \rightarrow 0^+}\!
	\oint_\gamma \frac{d\lambda}{4\pi^3 i}
	\ln(\frac{1+\epsilon+\lambda}{1+\epsilon-\lambda})
	\ln^2 \! \left( \frac{\sqrt{\lambda^2}+|\Theta|}{\sqrt{\lambda^2 -
	\Theta^2}} \right).
	\nonumber
\end{equation}
As for the tight-binding chain, $\gamma$ is the same dogbone contour of
Fig.~\ref{fig:contour}. After using~\eqref{eq:id-delta}, and proceeding
as for the tight-binding chain, we obtain
\begin{equation} \label{eq:kit-c}
    c'(\Theta) = \frac{3}{4\pi^2} \int_{-\Theta}^\Theta dx
    \ln(\frac{1+x}{1-x}) \ln(\frac{\Theta + x}{\Theta - x}).
\end{equation}
Remarkably, Eq.~\eqref{eq:kit-c} is half of the result of
Eq.~\eqref{eq:ctheta-int} obtained for the tight-binding chain
[see also Eq.~\eqref{eq:ctheta-integral}].


\bibliography{bibliography}

\end{document}